\documentclass[showpacs,amsmath,amssymb,twocolumn,pra,superscriptaddress,notitlepage]{revtex4-1}

\usepackage{qcircuit}
\usepackage{amsmath,bm}
\usepackage[dvips]{graphicx}
\usepackage{amsmath,amssymb,amsthm,mathrsfs,amsfonts,dsfont}
\usepackage{braket}
\usepackage{bm}
\usepackage{enumerate}
\usepackage{color}
\usepackage{graphicx}
\usepackage{algorithm}
\usepackage{algorithmic}
\usepackage[justification=RaggedRight]{caption}
\usepackage[subrefformat=parens]{subcaption}
\newcommand{\boltzmannP}{
P(\sigmavector|\uvector)
}
\newcommand{\boltzmanndist}[1]{
\dfrac{1}{Z(\uvector)}
\exp \left [
-{#1}
\right ]
}
\newcommand{\classicalH}{
H(\sigmavector,\uvector)
}
\newcommand{\uvector}{
\mbox{\boldmath $u$}
}
\newcommand{\sigmavector}{
\mbox{\boldmath $\sigma$}
}
\newcommand{\sumsigma}{
\sum_{\sigmarange} 
}
\newcommand{\sigmarange}{
\sigmavector \in\{-1,1\}^N
}
\newcommand{\hamilcalc}[2]{
-\sum_{i<j} J_{ij} {#1}{#2}
-\sum_{i=1}^N h_i {#1}
}
\newcommand{\empiricalP}{
P_D(\sigmavector)
}
\newcommand{\averagedata}{
\dfrac{1}{D}\sum_{d=1}^D 
}
\newcommand{\sigmadata}[1]{
\mbox{\boldmath $\sigma$}^{(#1)}
}
\newcommand{\deriv}[2]{
\dfrac{\partial {#1}}{\partial {#2}}
}

\begin{document}

% --------------------  TITLE  --------------------

\title{Boltzmann machine learning with a variational quantum algorithm}

% ------------  AUTHORS AND AFFILIATIONS ----------
\author{Yuta Shingu}
\affiliation{Department  of  Physics,  Faculty  of  Science  Division  I,Tokyo  University  of  Science,  Shinjuku,  Tokyo  162-8601,  Japan.}
\affiliation{Device Technology Research Institute,  National  Institute  of  Advanced  Industrial  Science  and  Technology  (AIST),1-1-1  Umezono,  Tsukuba,  Ibaraki  305-8568,  Japan.}

\author{Yuya Seki}
\affiliation{Device Technology Research Institute,  National  Institute  of  Advanced  Industrial  Science  and  Technology  (AIST),1-1-1  Umezono,  Tsukuba,  Ibaraki  305-8568,  Japan.}

\author{Shohei Watabe}
\affiliation{Department  of  Physics,  Faculty  of  Science  Division  I,Tokyo  University  of  Science,  Shinjuku,  Tokyo  162-8601,  Japan.}

\author{Suguru Endo}
\affiliation{Department of Materials, University of Oxford, Parks Road, Oxford OX1 3PH, United Kingdom}
\affiliation{NTT Computer and Data Science laboratories, NTT corporation, Musashino, Tokyo 180-8585, Japan}

\author{Yuichiro Matsuzaki}\email{matsuzaki.yuichiro@aist.go.jp}
\affiliation{Device Technology Research Institute,  National  Institute  of  Advanced  Industrial  Science  and  Technology  (AIST),1-1-1  Umezono,  Tsukuba,  Ibaraki  305-8568,  Japan.}

\author{Shiro Kawabata}
\affiliation{Device Technology Research Institute,  National  Institute  of  Advanced  Industrial  Science  and  Technology  (AIST),1-1-1  Umezono,  Tsukuba,  Ibaraki  305-8568,  Japan.}

\author{Tetsuro Nikuni}\email{nikuni@rs.tus.ac.jp}
\affiliation{Department  of  Physics,  Faculty  of  Science  Division  I,Tokyo  University  of  Science,  Shinjuku,  Tokyo  162-8601,  Japan.}

\author{Hideaki Hakoshima}\email{hakoshima-hideaki@aist.go.jp}
\affiliation{Device Technology Research Institute,  National  Institute  of  Advanced  Industrial  Science  and  Technology  (AIST),1-1-1  Umezono,  Tsukuba,  Ibaraki  305-8568,  Japan.}

% --------------------  ABSTRACT  --------------------

\begin{abstract}
A Boltzmann machine is a powerful tool for modeling probability distributions that govern the training data.
A thermal equilibrium state is typically used for the Boltzmann machine learning to obtain a suitable probability distribution.  
The Boltzmann machine learning consists of calculating the
gradient of the loss function given in terms of the thermal average, which is the most time
consuming procedure.
Here, we propose a method to implement the Boltzmann machine learning by using Noisy Intermediate-Scale Quantum (NISQ) devices. 
We prepare an initial pure state that contains all possible computational basis states with the same amplitude, and apply a variational imaginary time simulation. 
Readout of the state after the evolution in the computational basis approximates the probability distribution of the thermal equilibrium state that is used for the Boltzmann machine learning.
We perform the numerical simulations of our scheme and confirm that the Boltzmann machine learning works well.
Our scheme leads to a significant step toward an efficient machine learning using quantum hardware.
\end{abstract}

\maketitle

\section{Introduction}
Developing efficient learning algorithms for Boltzmann machine (BM)~\cite{fahlman1983massively,ackley1985learning}
is an important issue of machine learning.
A Boltzmann machine is a parametric stochastic model for the statistical machine learning,
which aims to extract patterns from data sets.
A restricted Boltzmann machine (RBM)~\cite{smolensky1986information,nair2010rectified,hinton2012practical},
a variant of BM, is a versatile model to represent an unknown distribution behind a given data set
because a RBM can approximate any discrete distribution~\cite{le2008representational}.
A restricted Boltzmann machine has a variety of applications such as dimensionality reduction~\cite{hinton2006reducing},
collaborative filtering~\cite{salakhutdinov2007restricted}, classification~\cite{larochelle2008classification}, quantum simulation~\cite{melko2019restricted,jia2019quantum,hsieh2019unitary},
topic modelling~\cite{hinton2009replicated}, feature learning~\cite{coates2011analysis},
and deep learning~\cite{dahl2011context,hinton2006fast,salakhutdinov2009deep,salakhutdinov2012efficient};
thus developing efficient learning algorithms of BM has benefits for the applications of machine learning. %\textcolor{red}{~\cite{hinton2002training,carreira2005contrastive,bengio2009justifying,gabrie2015training,marlin2010inductive,yasuda2012composite,yasuda2011learning}}.
However, the learning process of the %\remove{original} \modyseki{(YS comment: I removed this because the learning process of RBM need the calculations of expectation values too)}
BM requires calculations of expectation values
with respect to the Gibbs state, which are in general computationally hard.
%\textcolor{red}{%
%The standard approach on classical computers is to approximate the expectation values with sample averages using the Markov-chain Monte Carlo (MCMC) method~\cite{landau2014guide}.
%Calculation of the averages using the MCMC method typically requires many samplings, leading to long computational time for the learning process.
%However, it has been shown that the averages calculated from a few steps of the Markov chain are sufficient for the learning of RBM in practice~\cite{hinton2002training,carreira2005contrastive}.
%This method is called contrastive divergence (CD) learning, and faster than the standard MCMC method.
%However, a drawback of the CD learning is that the approximation introduces a bias which can prevent the model from converging to the optimal parameters~\cite{sutskever2010convergence}.
%To reduce the bias, parallel tempering Monte Carlo is applied for the learning of RBM~\cite{cho2010parallel,desjardins2010tempered}.
%In particular, the authors of Ref.~\cite{cho2010parallel} has revealed that the method is as efficient as the CD learning
%in terms of the computational time.
%}
To mitigate the computational cost, many approximate algorithms on classical computers have been developed~\cite{hinton2002training,carreira2005contrastive,bengio2009justifying,gabrie2015training,marlin2010inductive,yasuda2012composite,yasuda2011learning}.

%In order to enhance the performance of machine learning tasks, quantum machine learning, that is machine learning by using quantum computers, has been proposed. 
Quantum machine learning has been proposed for enhancing the performance of machine learning tasks by using quantum computers.
There are many algorithms of quantum machine learning with fault-tolerant quantum computers, such as quantum support vector machine \cite{rebentrost2014quantum}, linear regression \cite{schuld2016prediction}, data fitting \cite{wiebe2012quantum}, and quantum principal component analysis \cite{lloyd2014quantum}. 
%\textcolor{green}{Also, we can use a fault-tolerant quantum computer for BM learning. The quantum counterpart of the Boltzmann distribution is a Gibbs state $e^{- \beta H}/ \mathrm{Tr}[e^{- \beta H}]$, where $H$ is a Hamiltonian including not only $\sigma_z$ but also $\sigma_x$ and $\sigma_y$ terms (here, $\sigma_x,\sigma_y,\sigma_z$ are Pauli matrices) and $\beta$ is inverse temperature, and quantum metropolis algorithm can be used for sampling the Gibbs state~\cite{temme2011quantum,moussa2019measurement}, which allows for BM learning on quantum computers. Such a type of BM learning is called quantum Boltzmann machine learning \cite{amin2018quantum}.}
%\textcolor{green}{While exponential speedup over classical machine learning is ensured for those algorithms, they necessitate quantum phase estimation algorithm and hence a deep quantum circuit. }
 In the Noisy Intermediate-Scale Quantum (NISQ) era~\cite{preskill2018quantum,doi:10.7566/JPSJ.90.032001,cerezo2020variational2}, great efforts have been devoted to develop quantum algorithms. Among them, variational quantum algorithms (VQAs) have been considered to be the first useful application on NISQ computers, because they only need shallow quantum circuits~\cite{peruzzo2014variational,li2017efficient,mcclean2016theory,yuan2019theory,endo2018variational}. In VQAs, trial wave functions are generated from parametrized shallow depth quantum circuits and parameters are optimized by classical computers. Parametrized quantum circuits as a quantum analogue of a classical neural network may enable us to implement machine learning tasks, e.g., quantum circuit learning for supervised learning \cite{mitarai2018quantum} and data-driven quantum circuit learning for generative modelling~\cite{benedetti2019generative}. As quantum circuits are used for generating the trial wave function, one can leverage exponentially increasing Hilbert space in the number of qubits, which may enhance the representability of the model significantly.
 %\textcolor{green}{Furthermore, quantum Boltzmann machine, which prepares a Gibbs state by leveraging thermofield-double technique combined with a variational imaginary time simulation~\cite{yuan2019theory}, was proposed \cite{zoufal2020variational}. Note that this technique needs two copies of quantum states, which is not necessarily ideal for NISQ devices because the number of qubits is quite restricted for them.}

%In principle, a VQA, the variational imaginary time simulation~\cite{yuan2019theory}, enable one to implement the BM learning
%The BM learning based on VQAs can be in principle realized.
%This is because it needs expectation values of a Gibbs state, which can be prepared by leveraging thermofield-double technique combined with the variational imaginary time simulation
The BM learning based on VQAs can be in principle realized; the
required Gibbs state can be prepared by leveraging the thermofield-double
technique combined with the variational imaginary time
simulation~\cite{yuan2019theory}.
%, and 
%we can in principle use 
%this state could be in principle used
%for the BM learning 
%Interestingly, 
%This is because the scheme described in~\cite{yuan2019theory} can prepare the Gibbs states even with quantum Hamiltonians.
%or even the quantum BM learning~\cite{amin2018quantum}.
%is so general that any Gibbs states for arbitrary Hamiltonians can be prepared. 
Although this naive technique can implement even the quantum BM learning~\cite{amin2018quantum}, this technique needs two copies of quantum states, which is not favorable for NISQ devices.
%\textcolor{green}{remove:which is
%not necessarily ideal for NISQ devices because the number of qubits is quite restricted for them. }

In this paper, %motivated by the recent development of the quantum machine learning with NISQ devices, 
we propose a new scheme to implement the BM learning by using NISQ devices based on the variational imaginary time simulation.
%\textcolor{green}{with fewer qubits.}
%\textcolor{green}{Our work is motivated by recent development of the quantum machine learning with NISQ device.} 
%  \textcolor{green}{remove:
%  A Gibbs state can be prepared by leveraging thermofielddouble technique combined with the variational imaginary time simulation [31], and we can in principle use
%this state for Boltzmann machine learning.
%  However,  This technique needs two copies of quantum states.} 
%  \textcolor{green}{remove, which is not necessarily ideal for NISQ devices because the number of qubits is quite restricted for them.}
%Instead, in order to reduce the number of the required qubits,
%\textcolor{red}{Alternatively,}
%the previous work \cite{zoufal2020variational}, 
We use a pure initial state that contains all possible computational basis states with the same amplitude (that is a separable state of $\ket{++\cdots +}$, where $\ket{+}=\frac{1}{\sqrt{2}}(\ket{0}+\ket{1})$ is an eigenstate of $\hat{\sigma}_x$), and perform the imaginary time evolution on the initial state to obtain the desired state.
By reading out the state in the computational basis after applying the variational imaginary time simulation algorithm, the probability distribution becomes the same as what is required for the BM learning. 
Our proposal focuses on special cases to
simulate the classical BM where only diagonal terms of its Hamiltonian are relevant. 
The possible advantage of our scheme is that we do not need two copies of quantum states unlike the scheme in~\cite{yuan2019theory}, and this reduces the number of qubits to implement the algorithm.

The rest of this paper is organized as follows. In section~\ref{sec:BML}, we explain the standard setup of BM learning. In section~\ref{sec:variational imaginary time method}, we give the outline of the variational imaginary time simulation. In section~\ref{sec: learning with imaginary time simulation}, we propose our algorithm of BM learning with the variational imaginary time simulation. In section~\ref{sec:results}, we show the numerical results of our algorithm.
In section~\ref{sec:comparison}, we compare our method with other schemes of BM learning. Finally, we conclude and summarize this paper in section~\ref{sec:conclusion}.

\section{\label{sec:BML}Boltzmann machine learning}
%A Boltzmann machine is 
%a type of generative models,
A Boltzmann machine is a generative model that generates binary data according to a certain probability distribution.
In this paper, we focus on the fully visible BM, where all the visible units are connected  to each other,
because such a concise model is suitable to demonstrate our scheme for small systems.
%However, it should be noted that our scheme is also applicable to the RBM~\cite{smolensky1986information,nair2010rectified,hinton2012practical}.
%We will describe generalization of our scheme to the learning of the RBM in Sec.~\ref{sec: learning with imaginary time simulation}.

%\modyseki{(YS comment: I added a line break here.)}
Let $\sigmavector \in \{-1, 1\}^N$ be a set of binary parameters where  $\sigma_i$
denotes $i$-th component and $N$ is the total number of the units.
First, we define the learning model $\boltzmannP$ of BM by the Boltzmann distribution:
%At first, we let $\sigmavector \in \{-1, 1\}^N$ and $\sigmavector$'s i-th component written to $\sigma_i$, where $N$ is the total number of units.
\begin{equation}
\label{boltzmann_dist}
\boltzmannP
=\boltzmanndist{
\classicalH
},
\end{equation}
%\begin{equation}
%\label{dist_function}
%Z(\uvector)=
%\sumsigma
%\exp\{
%-\classicalH
%\}
%\end{equation}
%
where $Z(\uvector)$ is the partition function, and $\classicalH$ denotes the Ising Hamiltonian:
\begin{equation}
\label{boltzmann_H}
\classicalH
=\hamilcalc{\sigma_i}{\sigma_j}.
\end{equation}
%
%\begin{equation}
%\label{boltzmann_param}
%\uvector
%=
%(J_{12}, \cdots, J_{N-1 N},h_1, \cdots, h_N)^T
%\end{equation}
Here, $\uvector$ is a vector of the parameters of the Hamiltonian $\uvector =(J_{12},J_{13}, \cdots, J_{N-1 N},h_1, \cdots, h_N)^T$.
%\textcolor{green}{As the energy (the value of $\classicalH$) decreases, the probability $\boltzmannP$ increases based on Eq.~(\ref{boltzmann_dist}). }
%A lower energy (the value of $\classicalH$) configuration of $\sigmavector$ has a larger probability $\boltzmannP$ based on Eq.~(\ref{boltzmann_dist}).
%For example, with $J_{ij}>0$, $(\sigma_i, \sigma_j)=(-1,-1)$ or $(1,1)$ gives a lower energy
%than $(\sigma_i, \sigma_j)=(-1,1)$ or $(1,-1)$. 
%Also, with $h_i > 0$ ($h_i < 0$), $\sigma_i=1$ ($\sigma_i=-1$) shows a lower energy than $\sigma_i=-1$ ($\sigma_i=1$). 
%The total probability is determined by the combination of these effects based on the component of the Hamiltonian.

%Next, we 
We define an empirical distribution $\empiricalP$ derived from a training data-set as follows:
\begin{equation}
\label{empirical_dist}
\empiricalP
=\averagedata
\delta(
\sigmavector
, \ \sigmadata{d}
), %\ 
%\delta(\mbox{\boldmath $x$}, \ \mbox{\boldmath $y$})=
%\begin{cases}
%1 & (\mbox{\boldmath $x$} = \mbox{\boldmath $y$}) \\
%0 & (\mbox{\boldmath $x$} \neq \mbox{\boldmath $y$})
%\end{cases}
\end{equation}
where $\delta(\mbox{\boldmath $x$}, \ \mbox{\boldmath $y$})$ denotes a Kronecker delta of $N$ dimensions, $D$ denotes the total number of the training data, and
$\sigmadata{d} \in \{-1, 1\}^N$ ($d=1, \cdots, D$) denotes $d$-th binary values of the training data.

The purpose of BM is to find optimal parameters $\uvector$ such that the distribution in Eq.~(\ref{empirical_dist}) can be well reproduced by $\boltzmannP$. 
For this purpose, we need to minimize the Kullback-Leibler divergence (KLD):
%we need to adopt a measure to quantify a distance between two probability distributions, and we minimize the distance between them by choosing the optimized parameters.
%Specifically, we adopt the Kullback-Leibler divergence (KLD) as a measure:
\begin{align}
\label{kullback}
{\rm KL}(P_D||P)
&=&
\sumsigma
\empiricalP
\log
\dfrac{
\empiricalP
}{
\boltzmannP
},
\end{align}
which is a measure to quantify a distance between two probability distributions.
The KLD is always greater than or equal to zero, ${\rm KL}(P_D||P) \geq 0$, and the KLD becomes zero if and only if two probability distributions are identical. 
%In order to 
%consider the minimization of the KLD over all possible $\uvector$, we can focus on the second term of Eq. (\ref{kullback}).
%{\color{green} We can rewrite the KLD as follows
%\begin{eqnarray}
%\label{log_likelihood}
%KL(P_D||P)&=&
%\sumsigma
%\empiricalP
%\log \empiricalP
%- L(\uvector)
%\nonumber \\
%L(\uvector)
%&=&
%\sumsigma
%\empiricalP
%\log
%\boltzmannP
%\nonumber \\
%&=&
%\averagedata
%\log
%P(
%\mbox{\boldmath $\sigma$}
%=\mbox{\boldmath $\sigma$}^{(i)}
%|\mbox{\boldmath $u$})
%\end{eqnarray}
%where  $L(\uvector)$ denotes a log-likelihood function.} 
Minimization of the KLD over the parameter $\uvector$ is equivalent to the maximum likelihood estimation as a function of $\uvector$.
We use the gradient method to minimize the KLD over $\uvector$ at 
the $s$-th update:
\begin{equation}
\label{gradient_method}
\uvector [s+1]
=\uvector [s]
-\eta \left.\left[\deriv{}{\uvector}{\rm KL}(P_D||P)\right]\right\rvert_{\uvector=\uvector [s]} , %\label{updateym}
\end{equation}
%\begin{equation}
%\uvector[s]
%=
%\left(J_{12}[s],\cdots, J_{N-1 N}[s], h_1[s],\cdots, h_N[s] \right)^T
%\nonumber
%\end{equation}
%\begin{equation}
%\deriv{ }{\uvector}
%=
%\left(\deriv{ }{J_{12}},\cdots, \deriv{ }{J_{N-1 N}}, \deriv{ }{h_1},\cdots, \deriv{ }{h_N} \right)^T
%\nonumber
%\end{equation}
where $\eta$ denotes a learning rate satisfying $0<\eta \ll 1$, %$s$ denotes the iteration step, 
$\uvector [s]=\left(J_{12}[s],\cdots, J_{N-1 N}[s], h_1[s],\cdots, h_N[s] \right)^T$ denotes the set of the
learning parameter at the $s$-th update, and
$\partial / \partial \uvector=\left(\partial/\partial J_{12},\cdots, \partial/\partial J_{N-1 N}, \partial/\partial h_1,\cdots, \partial/\partial h_N \right)^T$ denotes the nabla operator of the parameter $\uvector$.
%By substituting Eq.~(\ref{log_likelihood}) into $\partial L / \partial \uvector$, 
By rewriting the derivative in Eq.~(\ref{gradient_method}),
we obtain the following expression:
%\begin{subequations}
\begin{align}
%\label{deriv_log_likelihood}
\deriv{ }{\uvector}{\rm KL}(P_D||P)
%&=& 
%\deriv{ }{\uvector}
%\averagedata
%\log
%P(
%\mbox{\boldmath $\sigma$}
%=\mbox{\boldmath $\sigma$}^{(i)}
%|\mbox{\boldmath $u$})
%\nonumber \\
&=
-\left \langle
\deriv{\classicalH}{\uvector}
\right \rangle  \notag\\
&\hphantom{={}}
+\averagedata
\deriv{
H(\mbox{\boldmath $\sigma$}
=\mbox{\boldmath $\sigma$}^{(d)}
,\mbox{\boldmath $u$})
}
{\uvector}.
%\\
%\label{deriv_log_likelihood_Jkl}
%\deriv{L}{J_{kl}}
%&=&
%\sumsigma
%\boltzmannP
%(-\sigma_k \sigma_l)
%-\averagedata
%(-\sigma_k^{(i)} \sigma_l^{(i)})
%\nonumber \\\\
%\deriv{L}{h_{k}}
%&=&
%\sumsigma
%\boltzmannP
%(-\sigma_k)
%-\averagedata
%(-\sigma_k^{(i)})
\label{Eq:derivative_of_KL}
\end{align}
%\end{subequations}
Here, the first term is the expectation value associated with $\boltzmannP$,
 where $\langle f(\boldsymbol{\sigma})\rangle=\sum_{\mbox{\boldmath $\sigma$}}f(\boldsymbol{\sigma}) \boltzmannP$ denotes a thermal equilibrium average of $f(\boldsymbol{\sigma})$.
%In Eq.~(\ref{deriv_log_likelihood}), 
The second term is the arithmetic mean with respect to training data, 
such as $\averagedata
\deriv{
H(\mbox{\boldmath $\sigma$}
=\mbox{\boldmath $\sigma$}^{(d)}
,\mbox{\boldmath $u$})
}
{J_{12}}=-\sum_{\mbox{\boldmath $\sigma$}} \sigma_1\sigma_2 \empiricalP$. 
This does not depend on the distribution $\boltzmannP$ obtained by the learning algorithm, and can be calculated before the learning algorithm. %\textcolor{green}{The second term does not depend on $\uvector$, and} 
Therefore we only have to calculate the first term every time we update the parameter %\modyseki{[comment: in this paper, \uvector is a singular form, it contains multiple parameters though.]} 
$\uvector$. However, a brute force approach to calculate the first term needs $O(2^N)$ steps, which is not tractable. In this paper, we propose a method to use a quantum algorithm to calculate the thermal equilibrium average.

\section{\label{sec:variational imaginary time method}Variational imaginary time simulation}
\begin{figure*}[h!t]
\begin{align*}
\Qcircuit @C=1em @R=.7em {
\lstick{(\ket{0}+e^{i\phi}\ket{1})/\sqrt{2}}&\qw&\qw&\gate{X}&\ctrl{2}&\gate{X}&\qw&\qw&\ctrl{2}&\gate{H}& \meter\\
&&...&&&&...&\\
\lstick{\ket{\bar{0}}}&\gate{U_1}&\qw&\gate{U_{k-1}}&\gate{u_{k,p}}&\gate{U_{k}}&\qw&\gate{U_{j-1}}&\gate{u_{j,q}}&\qw&\qw\\
}
\end{align*}
(a)
\begin{align*}
\Qcircuit @C=1em @R=.7em {
\lstick{(\ket{0}+e^{i \phi}\ket{1})/\sqrt{2}}&\qw&\qw&\gate{X}&\ctrl{2}&\gate{X}&\qw&\qw&\ctrl{2}&\gate{H}& \meter\\
&&...&&&&...&\\
\lstick{\ket{\bar{0}}}&\gate{U_1}&\qw&\gate{U_{k-1}}&\gate{u_{k,i}}&\gate{U_{k}}&\qw&\gate{U_{N}}&\gate{P_j}&\qw&\qw\\
}
\end{align*}
(b)
\caption{Quantum circuits for calculating (a) $\Re(e^{i\phi}\bra{\bar{0}}\mathcal{\hat{U}}_{k,p}^\dag \mathcal{\hat{U}}_{j,q}\ket{\bar{0}})$ and (b) $\Re(e^{i\phi}\bra{\bar{0}} \mathcal{\hat{U}}_{k,i}^\dag \hat{P}_j \hat{U}\ket{\bar{0}})$ in order to obtain $M_{k,j}$ in Eq.~(\ref{eq:M}) and $C_{k}$ in Eq.~(\ref{eq:V}). 
The upper horizontal line (the lower line) represents the ancillary qubit (the qubits of the system). We prepare the initial state $(\ket{0}+e^{i \phi}\ket{1})/\sqrt{2}$.
$X$ and $H$ denote the Pauli $X$ gate and the Hadamard gate, and $U_k$ ($k=1,\cdots, N_{\mathrm{ p}}$) is a parametrized unitary gate constituting the variational circuit.
Our ansatz circuit is shown in Fig.~2.
The expectation values of $Z$-measurement on the ancillary qubit give (a) $\Re(e^{i\phi}\bra{\bar{0}}\mathcal{\hat{U}}_{k,p}^\dag \mathcal{\hat{U}}_{j,q}\ket{\bar{0}})$ and (b) $\Re(e^{i\phi}\bra{\bar{0}} \mathcal{\hat{U}}_{k,i}^\dag \hat{P}_j \hat{U}\ket{\bar{0}})$.
Therefore by choosing $\phi=0$, $\pi/2$, we can obtain both the real and imaginary part of Eqs.~(\ref{eq:M}) and (\ref{eq:V}). 
%When $\sigma_{k,i}$ is Hermitian, the $X$ gates acting on the ancilla qubit can be also omitted. 
}\label{Fig:circuitPrac}
\end{figure*}
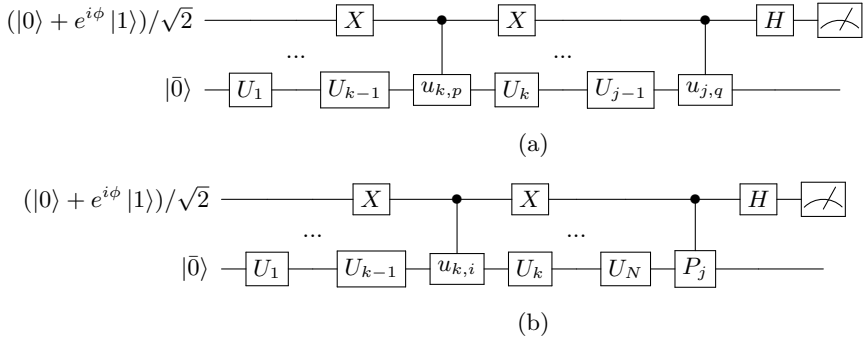
We here review the variational imaginary time simulation, which is compatible with NISQ devices. The Wick-rotated Schr\"odinger equation describing the imaginary time evolution can be written as  ~\cite{mcardle2019variational,yuan2019theory}: 
\begin{align}
\frac{d \ket{\psi(\tau)}}{d\tau}=-(\hat{H}-\bra{\psi(\tau)}\hat{H}\ket{\psi(\tau)}) \ket{\psi(\tau)}.
\label{Eq: wickrotated}
\end{align}
Here, this dynamics conserves the norm of $\ket{\psi(\tau)}$ due to the expectation value $\bra{\psi(\tau)}\hat{H}\ket{\psi(\tau)}$ in Eq.~(\ref{Eq: wickrotated}).
The state at time $\tau$ is expressed as:
\begin{align}
\label{state of wickrotated}
\ket{\psi(\tau)}= \frac{\mathrm{exp}(-\hat{H} \tau) \ket{\psi(0)}}{\sqrt{\bra{\psi(0)}\mathrm{exp}(-2 \hat{H} \tau) \ket{\psi(0)}}},
\end{align}
whose norm is unity.
%\textcolor{green}{Here,$\bra{\psi(\tau)}H\ket{\psi(\tau)}$.}

Instead of directly simulating the non-unitary imaginary time evolution Eq.~(\ref{Eq: wickrotated}), the variational imaginary time simulation algorithm \cite{mcardle2019variational} employs the parametrized wave function $\ket{\varphi (\vec{\theta}(\tau))}$ on a parametrized quantum circuit: 
\begin{equation}\label{eq:parametric quantum state}
\begin{aligned}
\ket{\varphi (\vec{\theta}(\tau))}&= \hat{U}(\vec{\theta}(\tau))\ket{\bar{0}}\\ \hat{U}(\vec{\theta})& =\hat{U}_{N_{\mathrm{ p}}}(\theta_{N_{\mathrm{ p}}})\dotsm \hat{U}_{2}(\theta_{2})\hat{U}_{1}(\theta_{1}),
\end{aligned}
\end{equation}
%\modyseki{[comment: 1.~Here and hereafter, $U$ needs hat symbol. 2.~The definition symbol $:=$ can be replaced with the usual equal symbol $=$. The definition symbol is used only here.]} 
where $\hat{U}_k(\theta_k)$ $(k=1,\dotsc ,N_{\mathrm{ p}})$ corresponds to a parametrized gate constituting the variational quantum circuit, $N_p$ is a total number of the parameters, $\theta_k$ is a real parameter and $\ket{\bar{0}}$ is an initial state of this ansatz which is usually chosen to be equal to $\ket{\psi(0)}$. 

Then, Eq.~(\ref{Eq: wickrotated}) is mapped onto the evolution of the parameters. Here, we use the McLachlan variational principle \cite{mclachlan1964variational,broeckhove1988equivalence} to derive the time derivative equations for $\vec{\theta}(\tau)$. We minimize the distance between the exact evolution and that of the parametrized trial state as
\begin{align}
\delta \left\|\Big{(}
\frac{\partial}{\partial \tau} + \hat{H}-\bra{\varphi (\vec{\theta}(\tau))}\hat{H}\ket{\varphi (\vec{\theta}(\tau))}
\Big{)}\ket{\varphi (\vec{\theta}(\tau))}  \right\|=0, 
\end{align}
% and we have
where the expectation value $\bra{\varphi (\vec{\theta}(\tau))}\hat{H}\ket{\varphi (\vec{\theta}(\tau))}
$ plays a role of normalizing the trial state $\ket{\varphi (\vec{\theta}(\tau))}$.
%which 
This variational principle leads to
\begin{align}
 M\frac{\partial \vec{\theta}(\tau)}{\partial \tau}=\vec{C},
 \label{Eq:MC}
\end{align}
where 
\begin{align}
M_{k,j}&=\Re\bigg( \frac{\partial \bra{\varphi(\vec{\theta}(\tau))}}{\partial \theta_k}\frac{\partial \ket{\varphi(\vec{\theta}(\tau))}}{\partial \theta_j} \bigg),\\
\label{Eq:Mk,j}
%\end{align}
%and 
%\begin{align}
C_k&= - \Re \bigg(\bra{\varphi(\vec{\theta}(\tau))}\hat{H} \frac{\partial \ket{\varphi(\vec{\theta}(\tau))}}{\partial \theta_k} \bigg).
\end{align}
Here, $\|\ket{\varphi} \|\equiv \braket{\varphi|\varphi}$, and $\Re(z)$ denotes the real part of $z$.
%and we assumed $\vec{\theta}(\tau)$ is a real parameter.  
%Solving Eq.~(\ref{Eq:MC}), %one can update the parameter. 
%for a small interval $\delta \tau$, the imaginary time evolution can be approximately %simulated when we use a suitable update rule
The size of $M$ and $\vec{C}$ does not depend on the number of the qubits, but depends on the number of the parameters, $N_{\mathrm{ p}}$.
The Euler method with a small parameter $\delta \tau$ can approximately simulate the imaginary time evolution (\ref{Eq:MC}), given by
\begin{equation}
    \vec{\theta}(\tau+\delta \tau)\simeq \vec{\theta}(\tau)+ M^{-1}(\tau)\cdot \vec{C}(\tau)\delta \tau.
\end{equation}
%where we use the Euler method.

Note that each element of $M$ and $\vec{C}$ can be efficiently evaluated on quantum circuits  shown in Fig.~\ref{Fig:circuitPrac}.
%, and we will confirm this in the below. 
The derivative of parametrized gates can generally be represented as follows:
\begin{eqnarray}
\frac{\partial \hat{U}_{k}}{\partial \theta_{k}} = \sum_{i} a_{k,i} \hat{U}_{k} \hat{u}_{k,i}.
\label{eq:expansion}
\end{eqnarray}
Here, $\hat{u}_{k,i}$ is a unitary operator, with $a_{k,i}$ being a complex coefficient. For example, %letting $\sigma_x$ be a Pauli X operator, for $U_k(\theta_k)=\mathrm{exp}(-i \theta_k \sigma_x)$, 
$d\hat{U}_k/d\theta_k = -i  \hat{U}_k(\theta_k)\hat{\sigma}_x$ for a single-qubit rotation $U_k(\theta_k)=\mathrm{exp}(-i \theta_k \hat{\sigma}_x)$, and one can see $a_{k,i}=-i$ and $\hat{u}_{k,i}=\hat{\sigma}_x$. Thus, the derivative of the parametrized state $\ket{\varphi(\vec{\theta}(\tau))}$ is
\begin{eqnarray}\label{Eq:partialstate}
\frac{\partial \ket{\varphi(\vec{\theta}(\tau))}}{\partial \theta_{k}} = \sum_{i} a_{k,i} \mathcal{\hat{U}}_{k,i} \ket{\bar{0}},
\end{eqnarray}
where %\modyseki{[comment: The symbol $\mathcal{U}$ needs hat symbol for consistency.]}
\begin{eqnarray}
\mathcal{\hat{U}}_{k,i}= \hat{U}_{N_{\mathrm{ p}}}  \cdots \hat{U}_{k+1} \hat{U}_{k} \hat{u}_{k,i} \hat{U}_{k-1} \cdots  \hat{U}_{1}.
\label{Eq: uki}
\end{eqnarray}
Using Eq.~(\ref{Eq: uki}), one can express $M_{k,j}$ as
\begin{eqnarray}
M_{k,j} = \sum_{p,q}\Re \left(
a^*_{k,p}a_{j,q} \bra{\bar{0}} \mathcal{\hat{U}}^\dag_{k,p} \mathcal{\hat{U}}_{j,q} \ket{\bar{0}}
\right).
\label{eq:M}
\end{eqnarray}
For $\vec{C}$, assuming that the Hamiltonian $\hat{H}$ can be decomposed as $\hat{H}=\sum_j f_j \hat{P}_j$, where $f_j$ is real and $\hat{P}_j$ is a tensor product of Pauli operators, we obtain
\begin{equation}\label{eq:V}
%\begin{aligned}
C_{k} = -\sum_{i,j} \Re \left(
a_{k,i} f_{j} \bra{\bar{0}} \hat{U}^\dag (\vec{\theta})\hat{P}_{j}\hat{\mathcal{U}}_{k,i} \ket{\bar{0}}
 \right).
%\end{aligned}
\end{equation}
%Notice that all the elements of $M$ and $\vec{C}$ have the form of
%\begin{equation*}
%b \Re \left( e^{i\phi} \bra{\bar{0}} \hat{V} \ket{\bar{0}} \right),
%\end{equation*}
%where $b$ and $\phi$ are real parameters determined by $a_{k,i}$, and $\hat{V}$ is a unitary operator, which is either $\mathcal{\hat{U}}^\dag_{k,p}\mathcal{\hat{U}}_{j,q}$ or $\hat{U}^\dag \hat{P}_{j}\mathcal{\hat{U}}_{k,i}$. 
It is known that one can calculate $M_{k,j}$ and $C_{k}$ by using the so-called Hadamard test with controlled unitary gates with the unitary operators of $\mathcal{\hat{U}}^\dag_{k,p} \mathcal{\hat{U}}_{j,q}$ and $\hat{U}^\dag (\vec{\theta})\hat{P}_{j}\hat{\mathcal{U}}_{k,i}$ \cite{nielsen2002quantum}.
However, such controlled unitary gates acting on many qubits require a large depth on the circuit, which may not be suitable for NISQ devices. Instead, to compute the elements of $M$ and $\vec{C}$, we can use a simplified quantum circuit as shown in  Fig.~\ref{Fig:circuitPrac}. Although this circuit contains the controlled unitary gates, only the control of $\hat{u}_{k,p}$ and $\hat{P}_j$ is required, and therefore the depth of this circuit is much shorter than that with the standard Hadamard test using controlled unitary gates $\mathcal{\hat{U}}^\dag_{k,p} \mathcal{\hat{U}}_{j,q}$ and $\hat{U}^\dag (\vec{\theta})\hat{P}_{j}\hat{\mathcal{U}}_{k,i} $.
The detailed calculation of this simplified Hadamard test in Fig.~\ref{Fig:circuitPrac} is explained in Appendix A.

The necessary number of the circuits to evaluate the matrix $M$ is equivalent to the independent elements of the matrix.
When we evaluate the vector $C$, we need to decompose the Hamiltonian and it needs more circuits proportional to the number of the terms in the Hamiltonian.
These elements are determined by the total number of the parameters $N_{\mathrm{ p}}$.
Note that the matrix $M$ is a Hermitian $N_{\mathrm{p}}\times N_{\mathrm{p}}$ matrix and the vector $\vec{C}$ is a $N_{\mathrm{p}}$ dimensional vector, and so the total number of the independent components is $O(N_{\mathrm{p}}^2)$.

In Appendix B, we discuss the effect of the shot noise in obtaining the elements of the matrix $M$ and the vector $\vec{C}$ with an accuracy $\varepsilon_s$.
We show an important relation between the KLD and the fidelity (or the trace distance) for sufficiently small $\varepsilon_s$, and we give a derivation of the condition for the required number of measurements $N_m$~\cite{li2017efficient,endo2019hybrid,PhysRevResearch.2.033281}.
In order to obtain the expectation values within the deviation $\varepsilon_s$, the required number of measurements is $N_m=O(T^2/\varepsilon_s^2)$ on the time interval $0\le \tau \le T$. %We also show that in the same condition of $N_m$, the KLD and the fidelity between the obtained state and the true state are less than $\varepsilon_s$.
If this condition is satisfied by taking a large number of measurements, we obtain both the KLD and the trace distance within the accuracy $\varepsilon_s$, except for a constant factor.

%The necessary number of measurements per circuit is shown in Appendix D, in order to obtain the elements of the matrix $M$ and the vector $\vec{C}$ with an accuracy $\varepsilon_s$.

%\textcolor{green}{According to the usual procedure, one can calculate the expectation values of $M_{k,j}$ and $C_{k}$ by using the Hadamard test with controlled unitary gates (the unitary operators are $\mathcal{\hat{U}}^\dag_{k,p} \mathcal{\hat{U}}_{j,q}$ and $\hat{U}^\dag (\vec{\theta})\hat{P}_{j}\hat{\mathcal{U}}_{k,i} $).
%Instead of such fully controlled unitary gates, one can compute elements of $M$ and $\vec{C}$ by leveraging  quantum circuits shown in Fig.~\ref{Fig:circuitPrac}. 
%Although this quantum circuits are also kinds of the Hamadard test, the controlled unitary operators only needs controlled $u_{k,p}$ and $P_j$, and therefore this realizes more shallow circuits than that of the case by using fully controlled unitary operators.
%The details of this Hadamard test in Fig.~\ref{Fig:circuitPrac} are shown in Appendix C.} 

The previous result \cite{mitarai2019methodology} showed that the indirect measurements via the ancillary qubits in Fig.~\ref{Fig:circuitPrac} can be replaced by the direct measurements on the qubits of the system. %, which reduces the depth of the circuit.
This method mainly has two advantages.
First, one can reduce the depth of the quantum circuit.
Second, it allows one to replace the measurement on the ancillary qubits with the direct measurements of the system, which remove the use of the long-range gates between the ancillary qubit and the other qubits.
Both these two advantages are favorable for current NISQ devices.
%It is worth mentioning that we could reduce the depth of the quantum circuit in Fig.~\ref{Fig:circuitPrac} as follows.
%The role of the ancillary qubit in Fig.~\ref{Fig:circuitPrac} is an indirect measurement of the qubits of the system via the controlled unitary operations.
%In Ref.~\cite{mitarai2019methodology}, it is shown that such indirect measurements via the ancillary qubits can be replaced by the direct measurements on the qubits of the system, which reduces the depth of the circuit. 
%This approach has another advantage. 
%The quantum circuit in Fig.~\ref{Fig:circuitPrac} contains two-qubit gate operations between the ancillary qubit and the other qubits, which may need long range interactions.
%Although some physical systems such as ion traps and Rydberg atoms can realize long-range interactions \cite{zhang2017observation,bernien2017probing,richerme2014non,joshi2020quantum}, many other systems do not have such long range interactions. 
%The methods in \cite{mitarai2019methodology} allow us to replace the measurement on the ancillary qubits with the direct measurements of the system, which remove the use of the long-range gates. 
However, since the realized dynamics by using the method in \cite{mitarai2019methodology}  is the same as the dynamics realized by  Fig.~\ref{Fig:circuitPrac} under an assumption of no noise, we adopt the quantum circuit  Fig.~\ref{Fig:circuitPrac} in this paper.

\section{\label{sec: learning with imaginary time simulation}BM learning with variational imaginary time simulation}
In this section, we describe our proposal of
%We discuss how we can use 
a quantum algorithm to calculate the thermal equilibrium average in Eq.~(\ref{Eq:derivative_of_KL}). 
In particular, we adopt the variational imaginary time simulation described in the previous section. 
For this purpose, we replace $\sigma_i$ in the Hamiltonian (\ref{boltzmann_H}) with the Pauli matrix $\hat{\sigma}_{z,i}$.
We define $\ket{0}$ and $\ket{1}$ as the eigenstates of $\hat{\sigma}_z$ given by $\hat{ \sigma}_z \ket{0}=\ket{0}$ and $\hat{\sigma}_z \ket{1}= -\ket{1}$.
%We define the eigenstates of $\hat{\sigma}_z$ by $\ket{0},\ket{1}$, which correspond to the eigenvalues $\pm 1$ as $\hat{\sigma}_z\ket{0}=\ket{0}, \hat{\sigma}_z\ket{1}=-\ket{1}$.
%\begin{equation}
%\hat{\sigma}^z
%=
%\left(
%%1 & 0 \\
%0 & -1 \\
%\end{array}
%\right)
%\end{equation}
%\begin{equation}
%\hat{\sigma}^z_i
%=\overbrace{I \otimes \cdots \otimes I}^{i-1} \otimes \hat{\sigma}^z \otimes 
%\overbrace{I \otimes \cdots \otimes I}^{N-i}
%\end{equation}
%where, $\otimes$ denotes a tensor product, $I$ denotes a 2 by 2 identity matrix, and $\hat{\sigma}^z_i$ denotes a $2^N$ by $2^N$ matrix. 
%Also, we define eigenstates of $\hat{\sigma}^z$ as follows.
%
%\begin{equation}
%\ket{0}
%=
%\left(
%\begin{array}{cc}
%1 \\
%0 \\
%\end{array}
%\right)
%,
%\ket{1}
%=
%\left(
%\begin{array}{cc}
%0 \\
%1 \\
%\end{array}
%\right)
%\end{equation}

We thus introduce the quantum Hamiltonian as follows:
\begin{equation}
\label{hamil_matrix}
\hat{H}(\uvector)
=\hamilcalc{\hat{\sigma}_{z,i}}{\hat{\sigma}_{z,j}}
\end{equation}
By using the variational imaginary time simulation, we can prepare 
a quantum state:
\begin{equation}
%\label{psi}
\ket{\psi_{\uvector}}
=\sqrt{\dfrac{2^N}{Z(\uvector)}}
e^{-\hat{H}(\uvector)/2}\ket{++\cdots +}, \label{targetstates}
\end{equation}
using the initial state $\ket{++\cdots +}=\ket{+}\otimes\ket{+}\otimes \cdots \otimes \ket{+}$.
%\begin{equation}
%\label{goal}
%\ket{\psi}
%=
%\sqrt{\dfrac{2^N}{Z(\uvector)}}
%\exp \left\{
%-\dfrac{\hat{H}(\uvector)}{2}
%\right\}
%\sum^{2^N-1}_{m=0} \ket{m}
%\ket{\psi(0)}
%\end{equation}
%where $\ket{\psi(0)}$ denotes the initial state described as follows.
%\begin{equation}
%\ket{\psi(0)}
%=
%\overbrace{\ket{+} \otimes \cdots \otimes \ket{+}}^{N}
%\end{equation}
%Here, $\ket{+}=(\ket{0} + \ket{1})/\sqrt{2}$ is the eigenstate of $\hat{\sigma}_x$, and $\ket{++\cdots +}$ denotes tensor products of $\ket{+}$, $\ket{++\cdots +}=\ket{+}\otimes\ket{+}\otimes \cdots \otimes \ket{+}$. 
Since $\partial\hat{H}(\uvector)/\partial \uvector$ consists only of the Pauli matrix $\hat {\sigma}_z$, only diagonal elements of the density matrix $\ket{\psi_{\uvector}}\bra{\psi_{\uvector}}$ are involved
in calculating
%\remove{$\Braket{\psi_{\uvector}|
%\deriv{\hat{H}(\uvector)}{\uvector}
%|\psi_{\uvector}}$}
$\Braket{\psi_{\uvector}|\partial\hat{H}(\uvector)/\partial\uvector|\psi_{\uvector}}$.
Also, every diagonal element of $\ket{\psi_{\uvector}}\bra{\psi_{\uvector}}$ is the same as that in the 
Gibbs state $e^{-\hat{H}(\uvector)}/Z(\uvector)(= \sum_{\sigma}\boltzmannP \ket{\mbox{\boldmath $\sigma$}}\bra{\mbox{\boldmath $\sigma$}})$. So, once Eq.~(\ref{targetstates}) is prepared, one can calculate the thermal equilibrium average in the right hand side of Eq.~(\ref{Eq:derivative_of_KL}) as follows:
\begin{equation}
\label{psi}
\left \langle
\deriv{\classicalH}{\uvector}
\right \rangle
=
\Braket{\psi_{\uvector}|
\deriv{\hat{H}(\uvector)}{\uvector}
|\psi_{\uvector}}.
\end{equation}
%\textcolor{green}{This is because $\ket{\psi_{\uvector}}=\sum_{\sigma}\sqrt{\boltzmannP} \ket{\mbox{\boldmath $\sigma$}}$ and the density operator corresponding to $\ket{\psi_{\uvector}}$ has all the same diagonal elements about the computational basis as those of the Gibbs state \textcolor{blue}{$e^{-\hat{H}(\uvector)}/Z(\uvector)$ $(= \sum_{\sigma}\boltzmannP \ket{\mbox{\boldmath $\sigma$}}\bra{\mbox{\boldmath $\sigma$}})$}, and
%$\partial\hat{H}(\uvector)/\partial \uvector$ has only $\hat{\sigma}_z$ terms.}
Note that both Eq.~(\ref{state of wickrotated}) and Eq.~(\ref{targetstates}) are normalized, and therefore when the initial state $\ket{\psi(0)}=\ket{++\cdots +}$ evolves from $\tau=0$ to $\tau=1/2$ following Eq.~(\ref{Eq: wickrotated}), two states are equal. This evolution is required for every update of the parameter $\uvector$ in the learning process.

% a Gibbs state can be prepared by leveraging thermofield-double technique combined with the variational imaginary time simulation~\cite{yuan2019theory} \textcolor{red}{where two copies of quantum states are required.}
 %We can in principle use this state for Boltzmann machine learning.
%s, of which the diagonal terms of its Hamiltonian are only relevant,

The above scheme can be also applied to the learning process of the RBM~\cite{smolensky1986information,nair2010rectified,hinton2012practical}.
Since the RBM has connections only between a hidden unit and a visible unit,
we can construct the Hamiltonian for the RBM by simply removing the interactions within the hidden unit and within the visible unit from Eq.~\eqref{hamil_matrix}.
The most time-consuming part of the learning process of the RBM is to calculate the thermal equilibrium average as well.
The average is given by the same equation~\eqref{psi}.
Consequently, we can use our scheme to update the parameters in the Hamiltonian for the RBM.

\begin{algorithm}[H]
\caption{Optimization of learning parameter}
\algsetup{indent=2em}
\label{alg}
\begin{algorithmic}
%\REQUIRE{$ $ \\
%$N_{step} \geq 0$ \\
%$0<\eta \ll 1$}
%\ENSURE{optimized $P(\sigmavector|\uvector)$}
\STATE Initialize $\uvector$ with the normal distribution
\FOR{$s=1, 2, \dotsc, N_{\text{step}}$}
%\STATE $\tau = 0$
\STATE $\tau \leftarrow 0$
%\STATE $\vec{\theta} = \vec{0}$
\STATE $\vec{\theta} \leftarrow \vec{0}$
%\WHILE{$\tau < 1/2$}
\REPEAT
%\STATE $M_{k,j} = \sum_{i,j}\Re \left(a^*_{k,p}a_{j,q} \bra{\bar{0}} U^\dag_{k,p} U_{j,q} \ket{\bar{0}} \right)$
\STATE Calculate matrix $M$ using Eq.~\eqref{eq:M}
%\STATE $C_{k} = -\sum_{i,j} \Re \left(
%a^*_{k,i} f_{j} \bra{\bar{0}} R^\dag_{k,i}P_{j}R \ket{\bar{0}}
% \right)$
\STATE Calculate vector $\vec{C}$ using Eq.~\eqref{eq:V}
%\STATE $\vec{\theta}(\tau + d\tau) = \vec{\theta}(\tau) + M^{-1} C d\tau$
\STATE $\vec{\theta} \leftarrow \vec{\theta} + M^{-1} \vec{C} \,\delta\tau$
%\STATE $\tau = \tau + d\tau$
\STATE $\tau \leftarrow \tau + \delta\tau$
\UNTIL $\tau = 1/2$
\STATE $\ket{\psi_{\uvector}} \leftarrow \ket{\varphi (\vec{\theta})}$ using Eq.~\eqref{eq:parametric quantum state}%
\STATE{$
\begin{aligned}
\deriv{ {\rm KL}}{\uvector}
&\leftarrow
-\Braket{\psi_{\uvector}|
\left.\deriv{\hat{H}(\uvector')}{\uvector'}\right\rvert_{\uvector' = \uvector}
|\psi_{\uvector}}\\
&\hphantom{\leftarrow{}}
+\averagedata
\left.\deriv{
H(\mbox{\boldmath $\sigma$}
=\mbox{\boldmath $\sigma$}^{(i)}
, \uvector')
}
{\uvector'}\right\rvert_{\uvector' = \uvector}
\end{aligned}
$
}
%\STATE $\uvector[s+1] = \uvector[s] + \eta \deriv{L}{\uvector}$
\STATE $\uvector \leftarrow \uvector - \eta \deriv{ {\rm KL}}{\uvector}$
\ENDFOR
\RETURN $\uvector$
\end{algorithmic}
\end{algorithm}

\section{\label{sec:results}results}
\subsection{A first check of our scheme}
\subsubsection{Datasets from true Boltzmann distribution and our algorithm}
In this section, we show the performance of our scheme by numerical simulations.
For this purpose, we need to generate a training data set.
First, we randomly generate $L$ sets of parameters %$\{\uvector^{(j)}\}_{j=1}^{L}$
 $\uvector^{*}_l$ ($l=1,2,\cdots,L$).
%where $L$ denotes the number of the parameters.
We use a normal distribution with a mean of $0$ and a variance of $1$, ${\mathcal N}(0,1)$, to generate each component of 
$\uvector^{*}_l$ for all $l$.
We call $\uvector^{*}_l$ true parameters.
Second, for a given $\uvector^{*}_l$, we generate training data 
$\sigmadata{i,l} \in \{-1, 1\}^N$ ($i=1, \cdots, D$)
based on a probability distribution of
$P(\boldsymbol{\sigma}|\uvector^{*}_l)$, which provides us with $L$ groups of the training data.
Third, we calculate $P^{(l)}_D(
\mbox{\boldmath $\sigma$})$ in Eq.~(\ref{empirical_dist}) %up to the $N_{\text{step}}$-th step 
by using each group of the training data.
In our numerical simulations,  we adopt $N=4$, $L=30$, and $D=1000$.

Our simulation method %with the number of iteration steps $N_{\text{step}}$,
%a fixed step size of the imaginary time evolution $\delta\tau$,
%and the learning rate $\eta$ ($0 < \eta \ll 1$) 
is summarized in Algorithm~\ref{alg}.
Here, we choose the number of iteration steps $N_{\text{step}}=100$, a fixed step size of the imaginary time evolution $\delta\tau = 0.1$, and the learning rate $\eta=0.1$ in the numerical simulations.
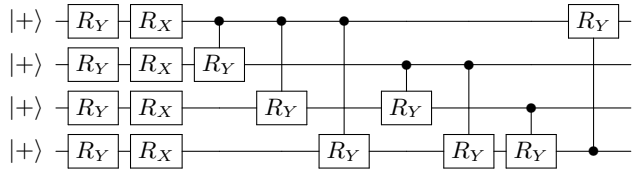
\begin{figure}[h!t]
\begin{align}
\Qcircuit @C=.5em @R=.4em {
\lstick{\ket{+}} & \gate{R_Y} & \gate{R_X} & \ctrl{1}   & \ctrl{2}   & \ctrl{3}   & \qw        & \qw        & \qw        & \gate{R_Y} & \qw \\
\lstick{\ket{+}} & \gate{R_Y} & \gate{R_X} & \gate{R_Y} & \qw        & \qw             & \ctrl{1}   & \ctrl{2}   & \qw        & \qw        & \qw \\
\lstick{\ket{+}} & \gate{R_Y} & \gate{R_X} & \qw        & \gate{R_Y} & \qw             & \gate{R_Y} & \qw        & \ctrl{1}   & \qw        & \qw \\
\lstick{\ket{+}} & \gate{R_Y} & \gate{R_X} & \qw        & \qw        &    \gate{R_Y} & \qw        & \gate{R_Y} & \gate{R_Y} & \ctrl{-3}  & \qw \\
}
\nonumber
\end{align}
\caption{Our ansatz circuit for the variational imaginary time simulation with 4 qubits.
We prepare the initial state $\ket{++++}$. $R_X$ and $R_Y$ represent rotational gates $R_X(\theta)=e^{-i\theta \hat{\sigma}_x/2}$ and $R_Y(\theta')=e^{-i\theta' \hat{\sigma}_y/2}$ with different parameter angles such as $\theta$ and $\theta'$.}
\label{fig:ansatz}
\end{figure}

Based on the training data, we run our algorithm as follows.
First, by using one of the groups of the training data generated by $P(\boldsymbol{\sigma}|\uvector_{1}^{*})$, 
we perform numerical simulations of our scheme where we choose our ansatz circuit for the variational imaginary time simulation as Fig.~\ref{fig:ansatz}, and obtain the learning parameters $\uvector_{\rm{opt}}^{(l)}$ after the optimization by using the gradient method Eq.~(\ref{gradient_method}) from $s=0$ to $N_{\text{step}}$. 
%It is worth mentioning that we 
For an initial guess of $\uvector$
for the simulation ($\uvector[0]$ in Eq.~(\ref{gradient_method})), we use the normal distribution with ${\mathcal N}(0,1)$ at each time.
Second, we then obtain $L$ sets of learning parameters after repeating the same numerical simulations but by using each group of the training data sampled from $P(\boldsymbol{\sigma} | \uvector^{*}_2)$, $P(\boldsymbol{\sigma}|\uvector^{*}_3)$, $\dots$, and $P(\boldsymbol{\sigma}|\uvector^{*}_L)$.
%we can perform 
%the same numerical simulations %\textcolor{green}{as the first step}
%but by using a different group of the training data sampled from $P(\boldsymbol{\sigma} | \uvector^{*}_2)$, $P(\boldsymbol{\sigma}|\uvector^{*}_3)$, $\dots$, and $P(\boldsymbol{\sigma}|\uvector^{*}_L$), we obtain a different learning parameter after the optimization. 
We repeat this process $L$ times.
Note that we set the number of each sampling to $10^4$ in order to obtain the expectation value in Eq.~(\ref{psi}) and to perform the variational imaginary time simulation using the quantum circuits in Fig.~\ref{Fig:circuitPrac}.

\subsubsection{Numerical simulations}
%\textcolor{green}{
%Firstly, we randomly generate $M$ sets of parameters $\uvector^{*}_m$ ($m=1,2,\cdots,M$).
%We use a normal distribution of a mean value of $0$ and a variance of $1$ to generate each component of 
%$\uvector^{*}_m$.
%Secondly, for \textcolor{green}{a given} $\uvector^{*}_m$, we generate $D$ \textcolor{green}{training data 
%$\sigmadata{i,m} \in \{-1, 1\}^N$} ($i=1, \cdots, D$)
%based on \textcolor{green}{the} probability distribution of $P(
%\mbox{\boldmath $\sigma$}
%|\mbox{\boldmath $u^{*}_m$})$, and we calculate \textcolor{green}{$P^{(m)}_D(
%\mbox{\boldmath $\sigma$})$ in Eq.~(\ref{empirical_dist}) with} each training data.
%Thirdly, %we divide the $D$ training data into $M$ groups.  
%Fourthly, 
%we perform a numerical simulations of our scheme by using one of the group of the training data, as summarized in Algorithm \ref{alg}. 
%Fourthly, we repeat the numerical simulations $M$ times
%by using different group of the training data.
%In our numerical simulation,  we adopt a case of $N=4$, 
%$D=1000$, and $M=30$. 
%It is worth mentioning that we need an initial guess of $\uvector$
%($\uvector[0]$ in the Eq. (\ref{gradient_method}))
%for the simulations, and we use the normal distribution of a mean value of $0$ and a variance of $1$ for the
%initial guess at each time.}

\begin{figure}
\centering
  \includegraphics[clip,width=0.5\textwidth]{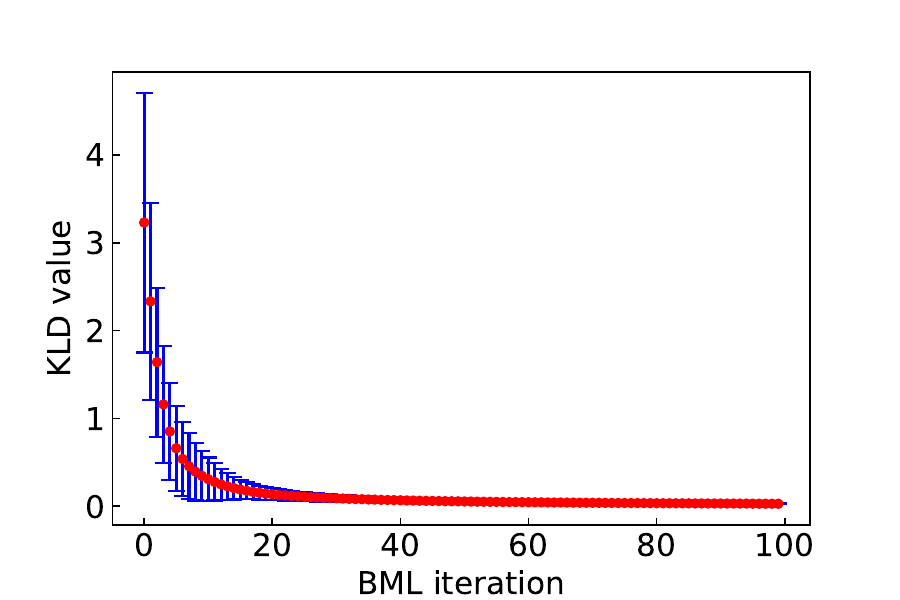}
   \caption{The KLD between the true distribution and the distribution obtained from the imaginary time evolution at each iteration step $s$.
   We set the total iteration step $N_{\text{step}}=100$ and plot the KLD from $s=0$ to $s=N_{\text{step}}$.
   We take an average of KLD over all groups of training data
   %for $m$ 
   at each step, and show each mean value and standard deviation in this plot.
   The KLD converges to zero, and thus our scheme 
  based on the imaginary time evolution can reproduce the training data.}
  \label{fig:kld}
\end{figure}

To evaluate the performance of our scheme, we consider the
%Fig.~\ref{fig:kld} shows the average 
KLD between the true distribution
%\remove{$P(\mbox{\boldmath $\sigma$}
%|\mbox{\boldmath $u^{*}_l$})$}
$P(\boldsymbol{\sigma}|\uvector^{*}_l)$
%$\uvector^{*}_l$
and 
estimated distribution
%the learning model
%\remove{$P(\mbox{\boldmath $\sigma$}
%|\mbox{\boldmath $u^{(l)}[s]$})$}
$P(\boldsymbol{\sigma}|\uvector^{(l)}[s])$
%and the learning model $\boltzmannP$ 
in Eq.~(\ref{boltzmann_dist}) 
optimized by the variational imaginary time simulation, 
where $\uvector^{(l)}[s]$ denotes a learning parameter at the $s$-th step.
Figure~\ref{fig:kld} shows the average and standard deviation of such KLD at $s$-th step
over all the true parameter sets $\uvector^{*}_l$ $(l=1,2,\dots, L)$.
%,
%and plot them 
%in Fig.~\ref{fig:kld}. 
These results show that all the average and the standard deviation of the KLD converge to zero.
Our scheme can thus successfully reproduce the probability distribution that has been generated the training data, and show the high performance of our proposal as 
BM learning.
%From these results, we can conclude that
 %almost all the KLD converge to zero in the same way and our proposal functions very well for Boltzmann machine learning.

\begin{figure}
\centering
  \includegraphics[clip,width=0.5\textwidth]{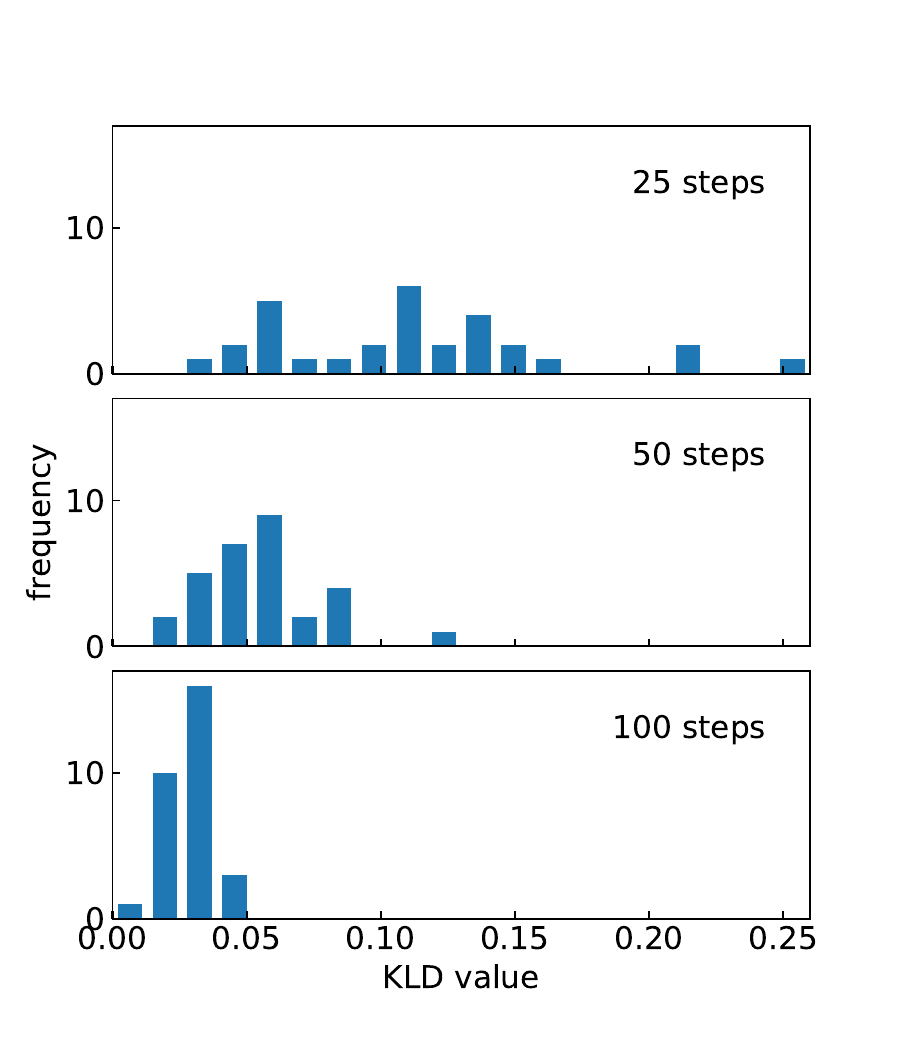}
  \caption{The histogram of the values of the KLD at $s=25$, $50$, and $100$ iteration step, respectively. We set the total iteration step $N_{\text{step}}=100$. 
  The range of the KLD values is from 0 to 0.26, and we divide it into 20 bins. We plot the histogram as the number of the counts in each bin.
  %Here, the histogram is a range of 0 to 0.26, divided into 20 segments, and if there's data in it, it counts 1.
  }
  \label{fig:hist}
\end{figure}

Furthermore, we investigate how the convergence speed of the KLD depends on the given true parameters $\uvector^{*}_l$.
Figure~\ref{fig:hist} shows the histogram of the values of the KLD at $s=$25, 50, and 100 iteration step, respectively.
At 25th iteration step, the values of the KLD strongly depends on the true parameters $\uvector^{*}_l$. However, as we increase the iteration step, the standard deviation becomes smaller, and 
most of the KLD becomes close to zero at 100th iteration step. 
%each value of the KLD varies and some KLDs converge fastly and others converge slowly. 
%From the graphs of 50th and 100th iteration steps, we can find all the values of the KLD approach zero.

\begin{figure}
  \includegraphics[width=0.47\textwidth]{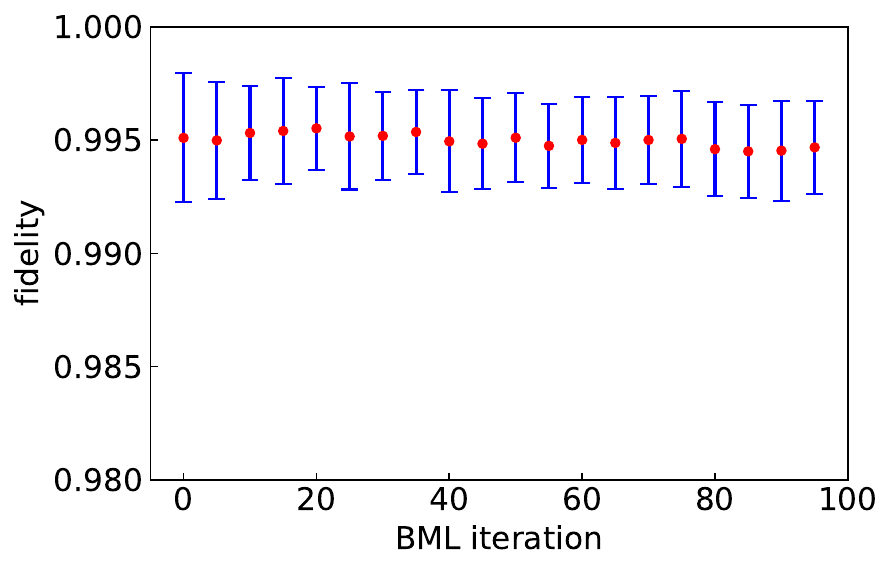}
  \caption{The fidelity between $\ket{\psi_{\uvector}}$ in  Eq.~(\ref{targetstates}) and $\ket{\varphi (\vec{\theta}(\tau =1/2))}$ in Eq.~(\ref{eq:parametric quantum state}) with respect to the iteration step $s$. The red points and the blue bars represent the average values and the standard deviation values of the fidelity over all groups of training data at each iteration step $s$.
  The fidelity is around or more than $0.995$ at every iteration step. These results show that our ansatz circuit is suitable to generate $\ket{\psi_{\uvector}}$ in Eq.~(\ref{targetstates}) with a reasonably good approximation.}
  \label{fig:fidelity}
\end{figure}

In addition, we examine whether the variational imaginary time simulation actually provides us with $\ket{\psi_{\uvector}}$ in Eq.~(\ref{targetstates}).
It is worth mentioning that 
%the imaginary time evolution does not always gives the desired state.
%Depending on the choise of the ansatz of the quantum circuit, there is a possibility that 
the state obtained by the variational imaginary time simulation $\ket{\varphi (\vec{\theta}(\tau =1/2))}$ in Eq.~(\ref{eq:parametric quantum state}) could be different from $\ket{\psi_{\uvector}}$ in  Eq.~(\ref{targetstates}), because its performance strongly depends on the ansatz of the quantum circuit. %Therefore, our machine learning algorithm may work well accidentally because we only use the expectation values in our scheme and not use the full information of the obtained state.
To investigate how close the state $\ket{\varphi (\vec{\theta}(\tau=1/2))}$ is to the state $\ket{\psi_{\uvector}}$, we calculate the fidelity:
\begin{align}
    F(\ket{\psi_{\uvector}},\ket{\varphi (\vec{\theta}(\tau=1/2))})=|\braket{\psi_{\uvector}|\varphi (\vec{\theta}(\tau=1/2))}|^2.
    \label{Eq:purefidelity}
\end{align}
The fidelity is more than $0.995$ at every iteration step (Fig.~\ref{fig:fidelity}).
%and we plot the fidelity between them in Fig.~\ref{fig:fidelity}. The fidelity is more than $0.995$ at every iteration step.
These results show the good agreement between $\ket{\psi_{\uvector}}$  and $\ket{\varphi (\vec{\theta}(\tau=1/2))}$.
From these results, we can exclude the possibility that our machine learning algorithm may work well accidentally, because we only use the expectation values in our scheme and not use the full information of the obtained state.
%Importantly, although the fidelity is not unity, our Boltzmann machine learning based in the imaginary time evolution still works to reproduce the training data.

\subsection{Depth number dependence}
In this subsection, we evaluate the required number of gates in order to achieve a sufficiently high fidelity between $\ket{\psi_{\uvector}}$ %in  Eq.~(\ref{targetstates}) 
and $\ket{\varphi (\vec{\theta}(\tau =1/2))}$.
%The definition of the fidelity is given by Eq.~(\ref{Eq:purefidelity}).
%in Eq.~(\ref{eq:parametric quantum state}).
%Except for the depth $N_{{\rm depth}}$ and the number of qubits $N$, 
Similar to Sec.~\ref{sec:results} A, 
we randomly generate $L$ sets of true parameters $\uvector^{*}_l$ ($l=1,2,\cdots,L$), and let
each component of $\uvector^{*}_l$ follow the normal distribution with ${\mathcal N}(0,1)$.

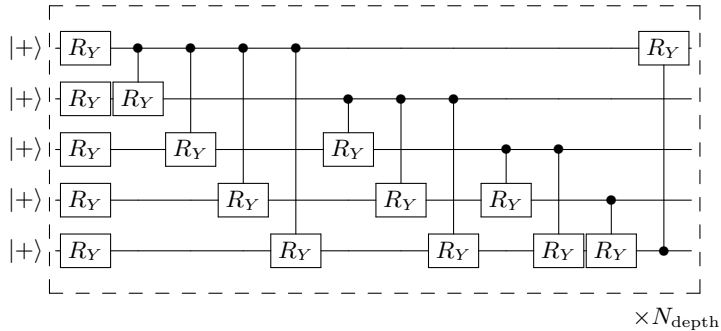
\begin{figure*}
\begin{align*}
\Qcircuit @C=.1em @R=.7em {
&&&&&&&&&&&&&&&&&&&&&&&\\
\lstick{\ket{+}}  &\qw&\gate{R_Y}&\ctrl{1}  &\ctrl{2}
&\ctrl{3}  &\ctrl{4}  &\qw       &\qw       &\qw       &\qw       &\qw
&\qw       &\gate{R_Y}&\qw\\
\lstick{\ket{+}}  &\qw&\gate{R_Y}&\gate{R_Y}&\qw
&\qw       &\qw       &\ctrl{1}  &\ctrl{2}  &\ctrl{3}  &\qw.      &\qw
&\qw       &\qw       &\qw\\
\lstick{\ket{+}}  &\qw&\gate{R_Y}&\qw       &\gate{R_Y}
&\qw       &\qw       &\gate{R_Y}&\qw       &\qw       &\ctrl{1}  &\ctrl{2}
&\qw       &\qw       &\qw\\
\lstick{\ket{+}}  &\qw&\gate{R_Y}&\qw       &\qw
&\gate{R_Y}&\qw       &\qw       &\gate{R_Y}&\qw       &\gate{R_Y}&\qw
&\ctrl{1}  &\qw       &\qw\\
\lstick{\ket{+}}  &\qw&\gate{R_Y}&\qw       &\qw
&\qw       &\gate{R_Y}&\qw       &\qw       &\gate{R_Y}&\qw       &\gate{R_Y}
&\gate{R_Y}&\ctrl{-4}  &\qw\\
&&&&&&&&&&&&&&&&&&&&&&\\
&&&&&&&&&&&\\
&&&&&&&&&&&&&&&&&&\hspace{-2em}\times N_{{\rm depth}}
\gategroup{1}{2}{7}{15}{0.7em}{--}
}
\nonumber
\end{align*}
\caption{Our ansatz circuit for the case of $N=5$ qubits. 
This is almost the same as Fig.~\ref{fig:ansatz} but repetitions.
$N_{{\rm depth}}$ denotes the number of repetitions.}
\label{fig:circuit depth}
\end{figure*}
In Fig.~\ref{fig:circuit depth}, we describe our ansatz circuit, where $N_{{\rm depth}}$ denotes the number of the repetitions. $N_{{\rm depth}}$ represents the power of the expression for our ansatz circuit. 
We perform the imaginary time evolution with this circuit, and investigate how the fidelity increases with $N_{{\rm depth}}$.  
We adopt parameters of $N=5, 6$, $L=100$, $N_{{\rm depth}}=1, 2, 3$, $\delta \tau=0.1$, and $10^4$ times measurements to obtain each matrix (vector) element of $M$ ($\vec{C}$).
%\textcolor{green}{remove:(
%In this section, we evaluate the efficiency of long-range interactions with respect to the fidelity between $\ket{\psi_{\uvector}}$ in  Eq.~(\ref{targetstates}) and $\ket{\varphi (\vec{\theta}(\tau =1/2))}$ in Eq.~(\ref{eq:parametric quantum state}). In the same way as the Section.~\ref{sec:results}, we randomly generate $L$ sets of true parameters $\uvector^{*}_l$ ($l=1,2,\cdots,L$). Each component of $\uvector^{*}_l$ follows the normal distribution which has a mean of 0 and a variance of 1. We define the number of the circuit depth $N_{depth}$ as shown Fig.~\ref{fig:circuit depth} and apply the circuit to our scheme. Also, we adopt parameters of $N=5, 6$, $L=100$, $N_{depth}=1, 2, 3$, $\delta \tau=0.1$.)}
\begin{figure*}
  \begin{minipage}[t]{0.5\linewidth}
   \centering
   \includegraphics[clip,width=0.9\textwidth]{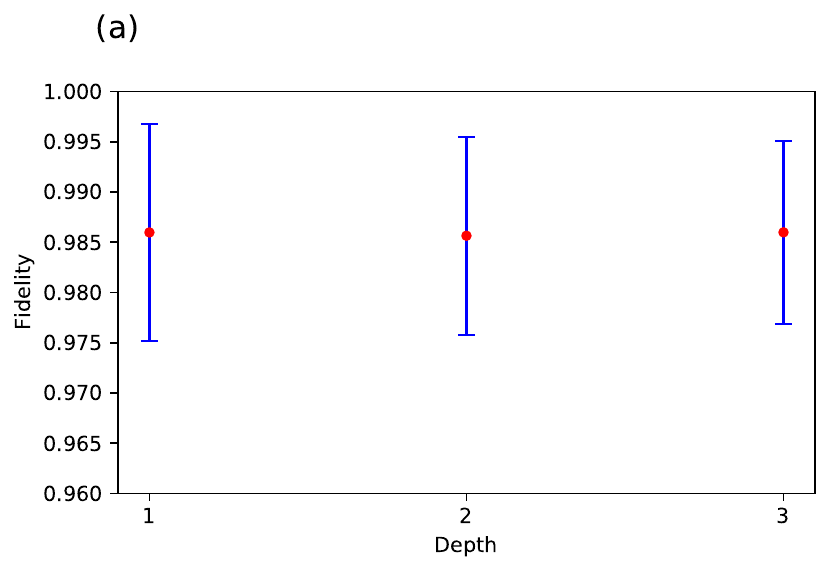}
%   \subcaption{
%   %Plot of the average fidelity in Eq.~(\ref{Eq:purefidelity}) against $N_{{\rm depth}}$ at 
%   $N=5$ qubits. }
  \label{fig:AppendixB fidelity q_num=5}
  \end{minipage}%
  \begin{minipage}[t]{0.5\linewidth}
    \centering
  \includegraphics[clip,width=0.9\textwidth]{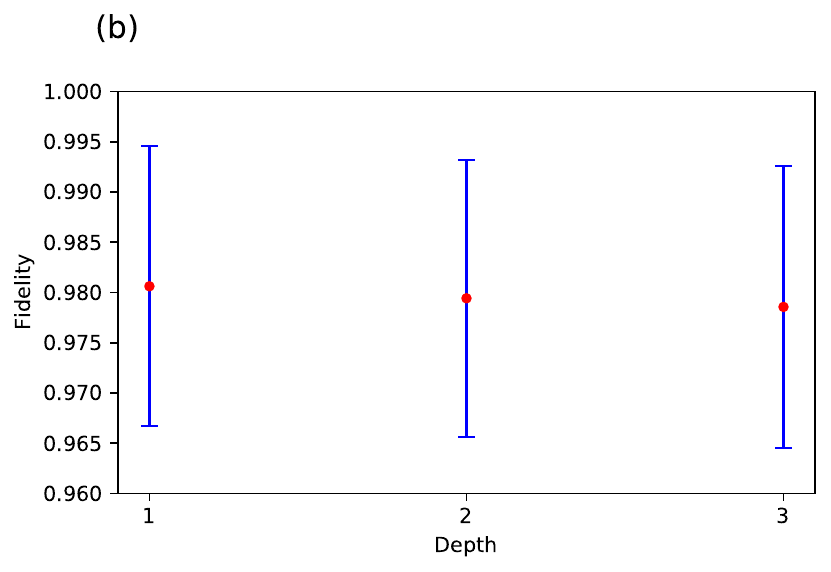}
%   \subcaption{$N=6$ qubits.
%   %Plot of the average fidelity in Eq.~(\ref{Eq:purefidelity}) against $N_{{\rm depth}}$ at $N=6$. The parameters are all the same as those in Fig.~\ref{fig:AppendixB fidelity q_num=5}.
%   }
  \label{fig:AppendixB fidelity q_num=6}
  \end{minipage}
  \caption{Plot of the average fidelity in Eq.~(\ref{Eq:purefidelity}) against $N_{{\rm depth}}$ at $N=5$ qubits in Fig.~7(a) and $N=6$ qubits in Fig.~7(b). We use the ansatz circuit in Fig.~\ref{fig:circuit depth} in order to calculate $\ket{\varphi (\vec{\theta}(\tau =1/2))}$ using the variational imaginary time simulation. The red points (the blue bars) denote the average values (the standard deviation) of the fidelity over all the true parameters $\uvector^*_l$ ($1\le l\le L$); we choose $L=100$.}
\end{figure*}
%\begin{figure}
%\centering
%  \includegraphics[clip,width=0.5\textwidth]{AppendixB_result_q_num5.pdf}
%   \caption{\textcolor{red}{Plot of the average fidelity in Eq.~(\ref{Eq:purefidelity}) against $N_{{\rm depth}}$ at $N=5$. We use the ansatz circuit in Fig.~\ref{fig:circuit depth} in order to calculate $\ket{\varphi (\vec{\theta}(\tau =1/2))}$ using the variational imaginary time simulation. The red points (the blue bars) denote the average values (the standard deviation) of the fidelity over all the true parameters $\uvector^*_l$ ($1\le l\le L$); we choose $L=100$.}}
%  \label{fig:AppendixB fidelity q_num=5}
%\end{figure}
%\begin{figure}
%\centering
%  \includegraphics[clip,width=0.5\textwidth]{AppendixB_result_q_num6.pdf}
%   \caption{\textcolor{red}{Plot of the average fidelity in Eq.~(\ref{Eq:purefidelity}) against $N_{{\rm depth}}$ at $N=6$. The parameters are all the same as those in Fig.~\ref{fig:AppendixB fidelity q_num=5}.}}
%  \label{fig:AppendixB fidelity q_num=6}
%\end{figure}

Figures~7(a)
%\ref{fig:AppendixB fidelity q_num=5} 
and~7(b)
%\ref{fig:AppendixB fidelity q_num=6}
show the average fidelity against $N_{{\rm depth}}$ at $N=5$ and $6$ for a given set of the true parameters $\uvector^*_l$ ($l=1,2,\cdots,L$), respectively. 
The definition of the fidelity is given by Eq.~(\ref{Eq:purefidelity}).
%The red points and the blue bars represent the average values and the standard deviation of the fidelity with respect to the true parameters $\uvector^*_l$ ($l=1,2,\cdots,L$) at each $N_{{\rm depth}}$.
%The key point of Fig.~\ref{fig:AppendixB fidelity q_num=5}(Fig.~\ref{fig:AppendixB fidelity q_num=6}) is 
Importantly, the values of the fidelity is around $0.985$($0.98$) for $N=5$ ($N=6$), even with $N_{{\rm depth}}=1$ and do not have a clear dependence on $N_{{\rm depth}}$. 
These results show that our ansatz circuit is suitable for our purpose even with $N_{{\rm depth}}=1$.
%These results show that the circuit which includes long-range interactions has the high performance to represent $\ket{\psi_{\uvector}}$ given by various parameters. Additionally, the circuit works well enough at $N_{depth}=1$.

\subsection{More general dataset}
In order to check the validity of our scheme, we investigate whether our scheme can fit a data set generated from a bars and stripes (BAS) \cite{Li_2020}, which cannot be exactly described by the Boltzmann distribution.
The way to generate the 4-dimensional BAS dataset is shown below:
Firstly, we randomly choose one from the following matrices $\left\{
\left(
    \begin{array}{rrr}
      -1 & -1 \\
      -1 & -1 
    \end{array}
\right),
\left(
    \begin{array}{rrr}
      -1 & 1 \\
      -1 & 1 
    \end{array}
\right),
\left(
    \begin{array}{rrr}
      1 & -1 \\
      1 & -1 
    \end{array}
\right),
\left(
    \begin{array}{rrr}
      1 & 1 \\
      1 & 1 
    \end{array}
\right)
\right\}
  $ with a probability of $1/4$. Secondly, we convert the matrix into a 4-dimensional vector where
  a matrix 
$\left(
    \begin{array}{rrr}
      a & b \\
      c & d 
    \end{array}
\right)$
is stochastically converted to either 
$\left(
    \begin{array}{rrrr}
      a & c & b & d 
    \end{array}
\right)^T$
or
$\left(
    \begin{array}{rrrr}
      a & b & c & d 
    \end{array}
\right)^T$ with a probability of $1/2$
where $a, b,c, d$ denote $1$ or $-1$.
Finally, we can get the dataset to repeat this procedure as many times as needed.
The distribution generated by the BAS has a feature that specific data such as $\left(
    \begin{array}{rrrr}
      1 & 1 & 1 & -1 
    \end{array}
\right)^T$ does not appear at all. On the other hand, in the distribution of the BM, every data could appear with non-zero probability. In this sense, it is not straightforward whether our scheme based on the BM could reproduce the dataset of BAS or not.

To simulate our scheme, we generate training data $\sigmadata{i} \in \{-1, 1\}^N$ ($i=1, \cdots, D$) from BAS. We also set up $L$ sets of initial values of learning parameters $\uvector^{(l)}[0]$ ($l=1, \cdots, L$) from the standard multivariate normal distribution to 
show that the results do not depend on specific initial values.
%indicate that the result does not involve the initial values. 
We adopt parameters of $N=4, L=30, D=1000, N_{{\rm step}}=100, \eta=0.1,$ and $\delta\tau=0.1$. Also, we use the quantum circuit in Fig.~\ref{fig:ansatz}, which is the same as that described in Section~\ref{sec:results}.
\begin{figure}
  \includegraphics[width=0.5\textwidth]{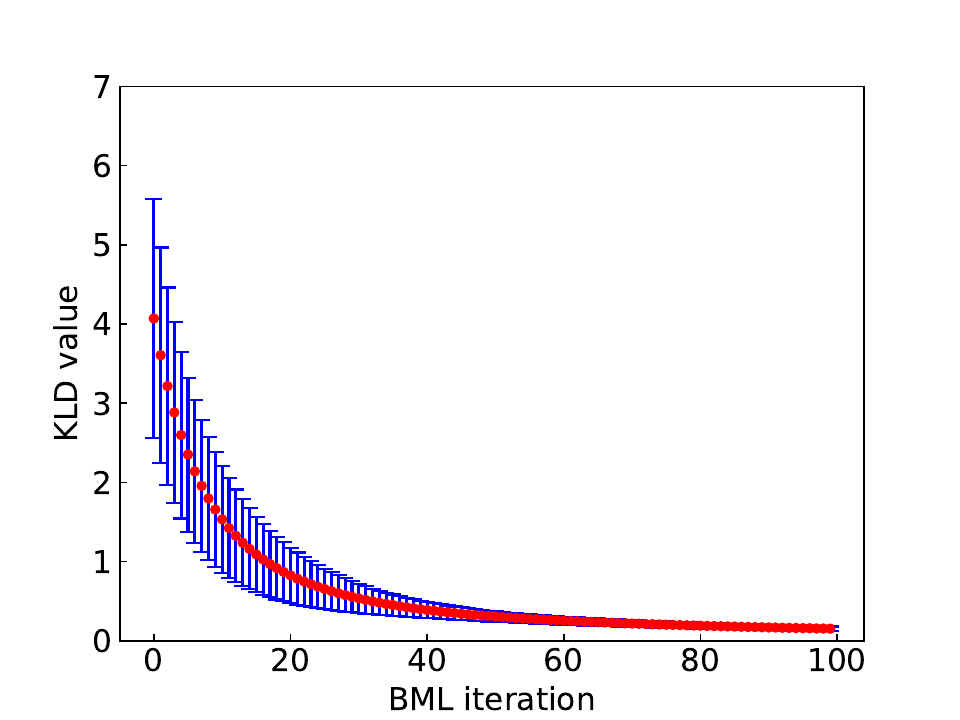}
   \caption{The KLD between the BAS distribution and the distribution obtained from the imaginary time evolution at each iteration step. We take an average of KLD over all groups of training data at each step, and
show  each  mean  value  and  standard  deviation  in  this  plot.}
  \label{fig:AppendixA kullback leibler}
\end{figure}

To evaluate the performance of our scheme toward the BAS dataset, we consider the KLD between the BAS distribution and the estimated distribution $P(\sigmavector \mid \uvector^{(l)}[s])$ in Eq.~(\ref{boltzmann_dist}) in the same way as Section~\ref{sec:results}. Figure~\ref{fig:AppendixA kullback leibler} shows the average and standard deviation of the KLD at $s$-th step over the learning parameters $\uvector^{(l)}[s]$ $(l=1,2, \cdots, L)$. The value of the KLD is around or less than 0.153 at 100th iteration step. So our simulation shows reasonable fittings with the given data.
\begin{figure}
  \includegraphics[width=0.5\textwidth]{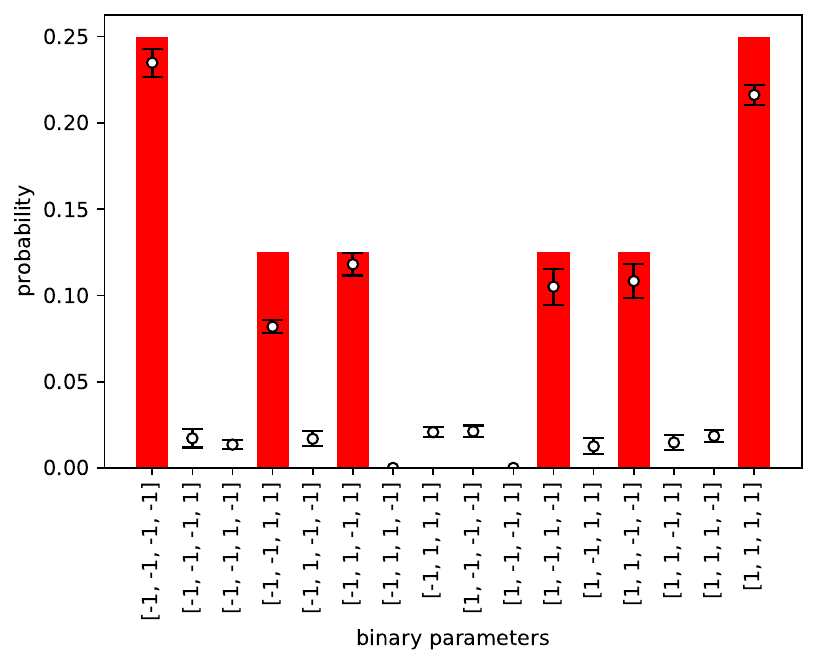}
   \caption{
   %Comparison of the BAS distribution with 
   Comparison between
   the BAS distribution and the distribution obtained from the 
   imaginary time evolution. Especially, for the imaginary time evolution, we 
   plot the average and standard deviation of the distribution
   %obtained by the imaginary time evolution
   at the 100th iteration step. We set the total iteration step $N_{{\rm step}}=100$.}
  \label{fig:AppendixA probability}
\end{figure}

Moreover, we compare the estimated distribution at 100th iteration step with the BAS distribution, and examine how the result varies with different initial parameters $\uvector^{(l)}[0]$.
%Moreover, we examine how the estimated distribution at 100th iteration step can represent the BAS distribution and relies on the given initial parameters $\uvector^{(l)}[0]$.
Figure~\ref{fig:AppendixA probability} shows the BAS distribution and the average and standard deviation of the estimated distributions at 100th step over the training parameters $\uvector^{(l)}[N_{{\rm step}}]$ $(l=1,2, \cdots, L)$. From this graph, we can confirm that the estimated distribution is capable of reproducing the BAS distribution, which is compatible with the results of the KLD in Fig.~\ref{fig:AppendixA kullback leibler}.
As described above, there remain finite values of the obtained distribution although the corresponding values are zero for the BAS distribution in Fig.~\ref{fig:AppendixA probability}. 
This deviation stems from the BM learning itself, not our scheme utilizing NISQ devices to obtain the required thermal average. %It is because our scheme is a method for obtaining a thermal average during the standard BM learning.
The learning performance can be improved by using the RBM learning with more ancillary qubits required in our scheme.
%as well as the case of Section.~\ref{sec:results}. 

%\textcolor{green}{In addition, we investigate whether our algorithm of the quantum imaginary time evolution works well.
%While the dynamics of the quantum imaginary time evolution is not unitary, the ansatz in the algorithm has only unitary operators.
%The simulation algorithm of the quantum imaginary time evolution gives the most optimal state in the ansatz that we choose.
%While only the expectation values are needed in Boltzmann machine, in Fig.~\ref{fig:fidelity}, we plot the fidelity between the target state (\ref{psi}) and the obtained state from the imaginary time evolution against the iteration steps.
%Here, we can rewrite $\ket{\psi_{\uvector}}$ in Eq.~(\ref{psi}) as
%\begin{equation}
%\label{target}
%\ket{\psi_{\uvector}}
%=\sum_{\sigma}\sqrt{\boltzmannP} \ket{\sigma},
%\end{equation}
%and we call Eq.~(\ref{target}) the target state here.
%From Fig.~\ref{fig:fidelity}, we can find that all the fidelity are near $1$ and the standard deviations of the fidelity are very small, so we can confirm that the quantum imaginary time evolution of our scheme also works well.}
%\textcolor{green}{Here, we choose our ansatz circuit for the imaginary time evolution as Fig.~\ref{fig:ansatz}.}

\section{\label{sec:comparison}Comparison with other schemes of Boltzmann machine learning}

Let us compare our scheme with other approaches for BM learning with classical and quantum algorithms.

The standard approach on classical computers is to approximate the expectation values with sample averages using the Markov-chain Monte Carlo (MCMC) method~\cite{landau2014guide}.
Calculation of the averages using the MCMC method typically requires many samplings, leading to long computational time for the learning process.
However, it has been shown that the averages calculated from a few steps of the Markov chain are sufficient for the learning of RBM in practice~\cite{hinton2002training,carreira2005contrastive}.
This method is called contrastive divergence (CD) learning, and faster than the standard MCMC method.
However, a drawback of the CD learning is that the approximation introduces a bias which can prevent the model from converging to the optimal parameters~\cite{sutskever2010convergence}.
To reduce the bias, parallel tempering Monte Carlo is applied for the learning of RBM~\cite{cho2010parallel,desjardins2010tempered}.
In particular, the authors of Ref.~\cite{cho2010parallel} has revealed that the method is as efficient as the CD learning
in terms of the computational time.

In addition to classical approaches,
there are other previous studies on the preparation of Eq.~(\ref{targetstates}) by using quantum annealing~\cite{somma2007quantum,yamamoto2020fair}. 
They construct the Hamiltonian whose ground state of which is equal to Eq.~(\ref{targetstates}). 
However, such a Hamiltonian includes many-body (more than three-body) interactions and is difficult to realize experimentally. 
Moreover, with the NISQ algorithms,
there is another method to create a thermal equilibrium state by minimizing the free energy function with a fixed temperature~\cite{wu2019variational,zhu2019variational,verdon2019quantum,chowdhury2020variational,wang2020variational,verdon2017quantum}.
Nevertheless, the free energy function is determined by the energy expectation value and the von Neumann entropy, and it is a non-trivial problem to calculate the von Neumann entropy on a relatively shallow quantum circuit.
%\textcolor{green}{Finally, only recently, the variational imaginary time simulation was used for quantum Boltzmann machine learning
%with numerical calculations and an IBM quantum device \cite{zoufal2020variational}. In their scheme, they can realize a thermal equilibrium with the Hamiltonian that contains not only $\hat{\sigma} _z$ terms but also
%$\hat{\sigma} _x$ and $\hat{\sigma} _y$ terms. 
%Instead, the necessary number of the qubits in their scheme is 
%A trade off is that their scheme requires to double the number of qubits with respect to the number of units in the original problem, which is twice as large as that of ours, which would show practical advantage of our scheme for implementing the BM learning by using just $\sigma _z$ terms.}

An interesting question for future work is whether our scheme is more useful due to the quantum properties or not, compared with the classical scheme of BM learning algorithms such as Markov-chain Monte Carlo samplings~\cite{landau2014guide}. In Appendix C, we give a detailed comparison with the CD learning for the case of a few qubits $N=4$. We show that our scheme exhibits rapid convergence. However,
the efficiency of the method may depend on various factors such as a data set and the number of qubits. Therefore, in order to answer the above question, we need a careful benchmark of the performance of our scheme by using an actual NISQ device, 
%that should contain at least hundreds of qubits, and compare it with the classical one. 
because it is difficult to obtain an analytical proof of quantum speedup as with many other NISQ algorithms. 
Actually, it is an open and central question whether a VQA has quantum advantage or not~\cite{cerezo2020variational2}.
%\textcolor{green}{(Hakoshima's comment: "difficulty" is not discussed in \cite{cerezo2020variational2}.) Actually, such a difficulty was discussed  in  a  recent  review  paper  of  NISQ  algorithms  \cite{cerezo2020variational2}.}
%, leaving this for future work.
However, without such an assessment about the quantum speedup, our work presented in this paper could still contribute to the society for the following two reasons.
First, most of the implementations of the BM learning are heuristic, and thus they may not always be able to find the suitable solution if we focus on just a specific implementation.
Therefore it is better if we have more options to perform the BM learning for searching the suitable solutions.  
Second, a quantum algorithm based on gate operations is compatible with a protocol called blind quantum computation~\cite{broadbent2009universal}. Here, a client can safely delegate a quantum algorithm to a server who has a gate-type quantum computer, and the server cannot steal any information about what the client performs with the server machine where the security is based on fundamental physics law such as no-signaling principle~\cite{morimae2013blind}. 
Since the training data could contain personal information that should not be leaked to the third party, such a secure BM learning could have a significant importance in the society.
Therefore, in principle, our scheme can be implemented with blind quantum computation, while information-theoretically secure BM learning is not known in classical implementations.
It is worth mentioning that such a secure quantum computer in the NISQ era was recently discussed in Ref.~\cite{kashefi2020securing}.

\section{\label{sec:conclusion}conclusion}
In conclusion, we propose a scheme to implement the BM learning based on the variational imaginary time simulation with NISQ devices. 
%Unlike 
%\textcolor{red}{While}
Our approach contrasts to the previous approach of Ref~\cite{yuan2019theory}, which prepares thermal equilibrium states
%The previous approaches 
%that
%\textcolor{red}{allow us to} prepare a
%thermal equilibrium state~\cite{yuan2019theory}
for quantum Hamiltonians by using two copies of quantum states. A key feature of our approach is to focus on classical cases when the Hamiltonian for the BM has only diagonal terms in the computational basis.
We use a pure state whose distribution mimics the thermal equilibrium distribution.
The potential advantage of our scheme
is that we do not need two copies of quantum states unlike the scheme of~\cite{yuan2019theory}, which reduces the number of qubits. 
%\textcolor{cyan}{Because we only simulate the classical systems, of which the diagonal terms of its Hamiltonian are only relevant, the} necessary number of the qubits of our scheme is twice smaller than that of the previous scheme to use the imaginary time evolution %to perform the BM learning. 
%to prepare the Gibbs state~\cite{yuan2019theory}.
Our results show potential for an efficient use of the NISQ device for the BM learning.

\begin{acknowledgments}
We thank Dr.~Muneki Yasuda for useful comments on BM.
We also thank Takashi Imoto and Atsuki Yoshinaga for valuable discussions.
This work was supported by Leading Initiative for Excellent Young Researchers MEXT Japan and JST presto (Grant No. JPMJPR1919) Japan. This work was partly supported by MEXT Q-LEAP (JPMXS0118068682), and JST ERATO (JPMJER1601).
This paper is partly based on results obtained from a project commissioned by the New Energy and Industrial Technology Development Organization (NEDO), Japan. 
S.W. was supported by Nanotech CUPAL, National Institute of Advanced Industrial Science and Technology (AIST).

We performed the numerical calculations in Figs.~\ref{fig:kld} and \ref{fig:hist} by using Qiskit, a open-source library for numerical simulations of quantum algorithms provided by Ref.~\cite{Qiskit}.

While this manuscript was being written, an independent article \cite{zoufal2020variational}  proposed quantum Boltzmann machine, which prepares a Gibbs state by leveraging thermofield-double technique combined with the variational imaginary time simulation (although this method for the Gibbs state preparation was already proposed in a published work~\cite{yuan2019theory}).

%While we are preparing our manuscript, we become aware
%of a related work that also uses variational imaginary time simulation for BM learning by .
%\textcolor{red}{copy and paste}
\end{acknowledgments}

\appendix

\section{Details of the Hadamard test}
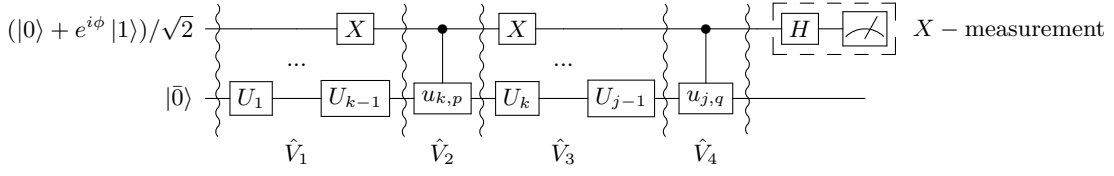
\begin{figure*}[h!t]
\begin{align*}
\Qcircuit @C=1em @R=.7em {
\lstick{(\ket{0}+e^{i\phi}\ket{1})/\sqrt{2}} 
\ar@{~}[]+<0.5em,1em>;[d]+<0.5em,-3em>
&\qw&\qw&\gate{X} 
\ar@{~}[]+<2em,1em>;[d]+<2em,-3em> &\ctrl{2} \ar@{~}[]+<1.6em,1em>;[d]+<1.6em,-3em>&\gate{X}&\qw&\qw
\ar@{~}[]+<1.8em,1em>;[d]+<1.8em,-3em> &\ctrl{2}
\ar@{~}[]+<1.7em,1em>;[]+<1.7em,-4.3em> &\qw&\gate{H}& \meter &\hspace{8em} X-{\rm measurement}\\
&&...&&&&...&\\
\lstick{\ket{\bar{0}}}&\gate{U_1}&\qw&\gate{U_{k-1}}&\gate{u_{k,p}}&\gate{U_{k}}&\qw&\gate{U_{j-1}}&\gate{u_{j,q}}&\qw&\qw&\qw \\
&&&&&&&&& \\
&&\hat{V}_1&&\hat{V}_2&&\hat{V}_3&&\hat{V}_4
\gategroup{1}{11}{1}{12}{0.7em}{--}
}
\nonumber
\end{align*}
\caption{Division of the unitary operation of the quantum circuit in Fig.~\ref{Fig:circuitPrac} (a) by $\hat{V}_1, \hat{V}_2, \hat{V}_3, \hat{V}_4$ in order to explain how to obtain the expectation values $\Re \bigg(
a^*_{k,p}a_{j,q} \bra{\bar{0}} \mathcal{\hat{U}}^\dag_{k,p} \mathcal{\hat{U}}_{j,q} \ket{\bar{0}}\bigg)$.}
\label{fig:AppendixC how to compute}
\end{figure*}
Here, we explain the details of obtaining $M_{k,j}$ by implementing the circuit in  Fig.~\ref{Fig:circuitPrac}.
%Let us comment why the expectation value 
% Eq.~\ref{Eq:Mk,j} 
%$M_{k,j}$ can be computable by running the circuit such Fig.~\ref{Fig:circuitPrac}.
Especially, for a given set of $k$, $p$, $j$, $q$, we show the detailed calculation of the term
%how to calculate one term 
$\Re \bigg(
a^*_{k,p}a_{j,q} \bra{\bar{0}} \mathcal{\hat{U}}^\dag_{k,p} \mathcal{\hat{U}}_{j,q} \ket{\bar{0}}\bigg)$, which corresponds to the right side of Eq.~(\ref{eq:M}).
Let us define two real variables $B$ and $\phi$ by %$B=|a^*_{k,p}a_{j,q}|$ and $\phi$ 
%as satisfying 
$a^*_{k,p}a_{j,q}=Be^{i\phi}$, and also $\hat{T}_{i,j}=\hat{U}_{j} \hat{U}_{j-1}\cdots \hat{U}_{i+1} \hat{U}_{i}$ ($i\le j$), which can simplify the form as $\mathcal{\hat{U}}_{k,i}=\hat{T}_{k,N_p}
 \hat{u}_{k,i} \hat{T}_{1,k-1}$.
As shown in Fig.~\ref{fig:AppendixC how to compute}, we separately consider the several gate operations (which we classify as $\hat{V}_1, \hat{V}_2, \hat{V}_3, \hat{V}_4$) of Fig.~\ref{Fig:circuitPrac}, 
%separate 
  %and 
  the $X$-measurement that consists of the Hadamard gate, 
  and the $Z$-measurement on the ancillary qubit.  
  Firstly, let us explain the changes in the quantum state via the implementation of the several gates.
  %show the flow of the circuit implementation as follows.
\begin{align}
    &\dfrac{1}{\sqrt{2}}(\ket{0}+e^{i\phi}\ket{1})\otimes\ket{\bar{0}} \notag\\
    \xrightarrow{\hat{V}_1}&\dfrac{1}{\sqrt{2}}(\ket{1}+e^{i\phi}\ket{0})\otimes \hat{T}_{1,k-1}\ket{\bar{0}} \notag\\
    \xrightarrow{\hat{V}_2}&\dfrac{1}{\sqrt{2}}\bigg(\ket{1}\otimes \hat{u}_{k,p}\hat{T}_{1,k-1}\ket{\bar{0}}
    +e^{i\phi}\ket{0}\otimes \hat{T}_{1,k-1}\ket{\bar{0}}\bigg) \notag\\
    \xrightarrow{\hat{V}_3}&\dfrac{1}{\sqrt{2}}\bigg(\ket{0}\otimes \hat{T}_{k,j-1} \hat{u}_{k,p}\hat{T}_{1,k-1}\ket{\bar{0}}+e^{i\phi}\ket{1}\otimes \hat{T}_{1,j-1}\ket{\bar{0}}\bigg) \notag\\
    \xrightarrow{\hat{V}_4}&\dfrac{1}{\sqrt{2}}\bigg(\ket{0}\otimes \hat{T}_{k,j-1} \hat{u}_{k,p}\hat{T}_{1,k-1}\ket{\bar{0}}+e^{i\phi}\ket{1}\otimes \hat{u}_{j,q}\hat{T}_{1,j-1}\ket{\bar{0}}\bigg) \notag\\
    =&\ket{\Phi_{k,j,p,q}}
\end{align}
where $\ket{\Phi_{k,j,p,q}}$ denotes $\hat{V}_4\hat{V}_3\hat{V}_2\hat{V}_1\bigg(\dfrac{1}{\sqrt{2}}(\ket{0}+e^{i\phi}\ket{1})\otimes\ket{\bar{0}}\bigg)$.
Secondly, let us explain how the $X$-measurement on the ancillary qubit provides the value of
%implementing the $X$-measurement and repeating to run the circuit as many times as needed, we obtain 
$\Re \bigg(
e^{i\phi} \bra{\bar{0}} \mathcal{\hat{U}}^\dag_{k,p} \mathcal{\hat{U}}_{j,q} \ket{\bar{0}}\bigg)$ as follows:
%\textcolor{blue}{
\begin{align}
    &\bra{\Phi_{k,j,p,q}}X\otimes I\ket{\Phi_{k,j,p,q}} \notag\\
    %=&
    %\dfrac{1}{2}\bigg(\bra{0}\otimes \bra{\bar{0}}T_{1,k-1}^\dag u_{k,p}^\dag T_{k,j-1}^\dag+e^{-i\phi}\bra{1}\otimes \bra{\bar{0}}T_{1,j-1}^\dag u_{j,q}^\dag \bigg) \notag \\
    %&X\otimes I\bigg(\ket{0}\otimes T_{k,j-1}u_{k,p}T_{1,k-1}\ket{\bar{0}}
    %+e^{i\phi}\ket{1}\otimes u_{j,q}T_{1,j-1}\ket{\bar{0}}\bigg) \notag\\
    =&
    \dfrac{1}{2}\bigg(\bra{1}\otimes \bra{\bar{0}}\hat{T}_{1,k-1}^\dag \hat{u}_{k,p}^\dag \hat{T}_{k,j-1}^\dag+e^{-i\phi}\bra{0}\otimes \bra{\bar{0}}\hat{T}_{1,j-1}^\dag \hat{u}_{j,q}^\dag \bigg) \notag \\
    &\bigg(\ket{0}\otimes \hat{T}_{k,j-1}\hat{u}_{k,p}\hat{T}_{1,k-1}\ket{\bar{0}}
    +e^{i\phi}\ket{1}\otimes \hat{u}_{j,q}\hat{T}_{1,j-1}\ket{\bar{0}}\bigg) \notag\\
    =&\dfrac{1}{2}\bigg(e^{-i\phi}\bra{\bar{0}}\hat{T}_{1,j-1}^\dag \hat{u}_{j,q}^\dag \hat{T}_{k,j-1}\hat{u}_{k,p}\hat{T}_{1,k-1}\ket{\bar{0}}\notag\\
    &+e^{i\phi}\bra{\bar{0}}\hat{T}_{1,k-1}^\dag \hat{u}_{k,p}^\dag \hat{T}_{k,j-1}^\dag \hat{u}_{j,q}\hat{T}_{1,j-1}\ket{\bar{0}}\bigg)\notag\\
    =&\dfrac{1}{2}\bigg(e^{-i\phi}\bra{\bar{0}} \mathcal{\hat{U}}^\dag_{j,q} \mathcal{\hat{U}}_{k,p} \ket{\bar{0}}+e^{i\phi}\bra{\bar{0}}\mathcal{\hat{U}}^\dag_{k,p} \mathcal{\hat{U}}_{j,q} \ket{\bar{0}}\bigg)\notag\\
    =&\Re \bigg(e^{i\phi}\bra{\bar{0}}\mathcal{\hat{U}}^\dag_{k,p} \mathcal{\hat{U}}_{j,q} \ket{\bar{0}}\bigg),
\end{align}
%}
where we apply $\hat{T}_{j,N_{\mathrm{ p}}}^\dag \hat{T}_{j,N_{\mathrm{ p}}}=I$ in the third line.
By multiplying $B$ by the expectation values of the measurement results, we obtain
%Thirdly, we obtain
$\Re \bigg(
Be^{i\phi} \bra{\bar{0}} \mathcal{\hat{U}}^\dag_{k,p} \mathcal{\hat{U}}_{j,q} \ket{\bar{0}}\bigg)$. 
%to multiply the circuit result by $B$.
Finally, by repeating the first and second step for every $p$ and $q$,
%repeating the first, second, and third step for all $p$ and $q$, then summing up the circuit results,
we obtain $M_{k,j}$ by summing up these results
as
\begin{align}
    M_{k,j} = \sum_{p,q}\Re \left(
Be^{i\phi} \bra{\bar{0}} \mathcal{\hat{U}}^\dag_{k,p} \mathcal{\hat{U}}_{j,q} \ket{\bar{0}}
\right).
\end{align}
We can also obtain $C_k$ in the same way.
\section{The effect of shot noise}
%As described in the main text, one can 
In this section, we consider the effect of noise when we compute elements of $M$ and $\vec{C}$ by utilizing  quantum circuits shown in Fig.~\ref{Fig:circuitPrac}.
It is worth mentioning that the effect of the decoherence and gate imperfections can be suppressed by quantum error mitigation technique \cite{temme2017error,endo2018practical,bonet2018low,endo2019mitigating,mcardle2019error,song2019quantum,sagastizabal2019experimental,kandala2019error,zhang2020error,hakoshima2020relationship,doi:10.7566/JPSJ.90.032001}.
Also, for simplicity, we assume that we choose the perfect ansatz to simulate the imaginary time evolution: 
\begin{align}
\ket{\psi_{\uvector}}&=\ket{\varphi (\vec{\theta}(\tau=1/2))}\\
\vec{\theta}(\tau=1/2)&=\int_0^{1/2} d\tau M_0 ^{-1}\vec{C}_0
\end{align}
where $M_0$ and $\vec{C}_0$ denote the ideal matrix of $M$ and $\vec{C}$ without shot noise.
In this case, we should mainly consider the shot noise that comes from the finite number of the repetitions of the measurements. 
So we investigate how many measurements are required
%how many numbers of measurements 
per circuit to obtain the accurate expectation values.
%and we neglect the other types of errors such as decoherence and the imperfection of the ansatz.
%\textcolor{green}{Please cite error mitigation papers here.}
%The argument of this section is based on Ref.~\cite{li2017efficient,endo2019hybrid,PhysRevResearch.2.033281}.

%Since we want to 
%We consider the effect of the shot noise.
The shot noise prevents us from performing a precise estimation of the $M$ and $\vec{C}$, and this leads a deviation of our ansatz quantum state (generated from the quantum circuit) from the ideal one.
We define the state $\ket{\psi_\varepsilon}$ with the shot noise as
\begin{align}
\ket{\psi_\varepsilon}&=\sqrt{1-\varepsilon}\ket{\psi_{\uvector}}+\sqrt{\varepsilon}\ket{\delta \psi_{\uvector}},
\end{align}
where $\ket{\psi_{\uvector}}$ is an ideal state without noise defined in Eq.~(\ref{targetstates})
%a noise-less state 
and $\ket{\delta \psi_{\uvector}}$ is a normalized state, $\|\ket{\delta \psi_{\uvector}}\|=1$ and $\braket{\psi_{\uvector}|\delta \psi_{\uvector}}=0$, which denotes the deviation of the ideal state due to
the shot noise.
%we choose $\rho_1=\ket{\psi_{\uvector}}\bra{\psi_{\uvector}}$ as a noise-less state and $\rho_2=\ket{\psi_\varepsilon}\bra{\psi_\varepsilon}$, where
%Here, we define the effect of the shot noise as $\sqrt{\varepsilon}\ket{\delta \psi_{\uvector}}$.
Here, we assume $\varepsilon$ is sufficiently small. 
%The Fuchs-van de Graaf's inequality between two states $\ket{\psi_{\uvector}}$ and $\ket{\psi_\varepsilon}$ is given by
Since we consider the pure states that are described by the wavefunctions
$\ket{\psi_{\uvector}}$ and
$\ket{\psi_\varepsilon}$, the following relation holds:
\begin{align}
\label{Eq:FuchvandeGraaf}
 D(\ket{\psi_{\uvector}},\ket{\psi_\varepsilon})&= \sqrt{1-F(\ket{\psi_{\uvector}},\ket{\psi_\varepsilon})}\\
F(\ket{\psi_{\uvector}},\ket{\psi_\varepsilon})&=|\braket{\psi_{\uvector}|\psi_\varepsilon}|^2=1-\varepsilon,
\end{align}
where $D(\ket{\psi_{\uvector}},\ket{\psi_\varepsilon})$ denotes the trace distance \cite{nielsen2002quantum}. 
%Since 
%these states 
%both
%$\ket{\psi_{\uvector}}$ and
%$\ket{\psi_\varepsilon}$ are pure states, this bound in Eq.~(\ref{Eq:FuchvandeGraaf}) is always achieved.
%\textcolor{green}{$\leftarrow$ YM cannot understand which bound is discussed here.
%}
%After calculating the fidelity between $\ket{\psi_{\uvector}}$ and $\ket{\psi_\varepsilon}$, we obtain $F(\ket{\psi_{\uvector}},\ket{\psi_\varepsilon})=1-\varepsilon$ and $\sqrt{1-F(\rho_1,\rho_2)}=\sqrt{\varepsilon}$.
We can give the bound of KLD between the true distribution function $\boltzmannP$ in Eq.~(\ref{boltzmann_dist}) and the distribution function obtained by the variational imaginary time simulation:
\begin{align}
P_\psi(\mbox{\boldmath $\sigma$} |\uvector)&=\frac{1}{2^N}|\braket{\mbox{\boldmath $\sigma$} |\psi_\varepsilon}|^2\\
&=\boltzmannP +2 \sqrt{\varepsilon \boltzmannP}\Re(\braket{\mbox{\boldmath $\sigma$} |\delta \psi_{\uvector}})+O(\varepsilon),
\end{align}
where $\ket{\mbox{\boldmath $\sigma$}}$ denotes the computational basis corresponding to $\mbox{\boldmath $\sigma$}$ in $\boltzmannP$.
We can calculate the KLD as
\begin{align}
{\rm KL}(P||P_\psi)&\simeq-2\sum_{\mbox{\boldmath $\sigma$}}\sqrt{\varepsilon \boltzmannP}\Re(\braket{\mbox{\boldmath $\sigma$} |\delta \psi_{\uvector}}),\notag\\
&\le 2 \sqrt{\varepsilon} \sqrt{\sum_{\mbox{\boldmath $\sigma$}}|\Re(\braket{\mbox{\boldmath $\sigma$} |\delta \psi_{\uvector}})|^2}\notag\\
&\le 2 \sqrt{\varepsilon} \sqrt{\sum_{\mbox{\boldmath $\sigma$}}|\braket{\mbox{\boldmath $\sigma$} |\delta \psi_{\uvector}}|^2}\notag\\
&= 2 \sqrt{\varepsilon}.
\end{align}
Therefore, we obtain
\begin{align}
{\rm KL}(P||P_\psi)\le 2\sqrt{1-F(\ket{\psi_{\uvector}},\ket{\psi_\varepsilon})}&= 2 D(\ket{\psi_{\uvector}},\ket{\psi_\varepsilon}).%\le \varepsilon_s
\label{Eq:Inequalities}
\end{align}
%and we can obtain the accuracy $\varepsilon_s$ by setting the number of measurements in  Eq.~(\ref{Eq:numberofmeasurements}).
Using this inequality, we will show the condition for the required number of measurements~\cite{li2017efficient,endo2019hybrid,PhysRevResearch.2.033281}.

The solution of Eq.~(\ref{Eq:MC}) can be rewritten as
\begin{align}
 \frac{\partial \vec{\theta}(\tau)}{\partial \tau}=(M_0+\delta M)^{-1}(\vec{C}_0+\delta \vec{C}),
 \label{Eq:imaginaryshotnoise}
\end{align}
where $\delta M$ and $\delta \vec{C}$ denote the deviation originating from shot noise.
If we assume that $\delta M$ is sufficiently small compared with $M_0$, Eq.~(\ref{Eq:imaginaryshotnoise}) can be rewritten as
\begin{align}
 \frac{\partial \vec{\theta}(\tau)}{\partial \tau}\simeq M_0^{-1}\vec{C}_0+M_0^{-1}\delta \vec{C}-M_0^{-2}\delta M\vec{C}_0,
 \label{Eq:imaginaryshotnoise2}
\end{align}
where we employ the linear approximation $(M_0+\delta M)^{-1}\simeq M_0^{-1}-M_0^{-2}\delta M$, neglecting the terms of higher order in  %and we neglect the higher order terms of 
$\delta M$ and $\delta \vec{C}$.
In Eq.~(\ref{Eq:imaginaryshotnoise2}), the first term of the RHS denotes the noise free solution and the second and the third terms represent the effect of the shot noise, and then we can evaluate the noise terms as
\begin{align}
\left|\delta\vec{\dot{\theta}}\right|
 =&\left|\frac{\partial \vec{\theta}(\tau)}{\partial \tau}-M_0^{-1}\vec{C}_0\right|\\\simeq& \left|M_0^{-1}\delta \vec{C}-M_0^{-2}\delta M\vec{C}_0\right|,\\
 \le &\left\|M_0^{-1}\right\| |\delta \vec{C}| +\left\|M_0^{-1}\right\|^2 \left\|\delta M\right\| |\vec{C}_0|,
 \label{Eq:imaginaryshotnoisenorm}
\end{align}
where $\|\cdots \|$ denotes the Frobenius norm.
From the central limit theorem, the noise terms behave as
\begin{align}
\left\|\delta M\right\|\simeq&\frac{\Delta_M}{\sqrt{N_m}},\\
 |\delta \vec{C}|\simeq&\frac{\Delta_{\vec{C}}}{\sqrt{N_m}},
 \label{Eq:imaginaryshotnoiseNm}
\end{align}
where $N_m$ denotes the number of measurements per quantum circuit.
Here, $\Delta_M$ and $\Delta_{\vec{C}}$ can be evaluated as
\begin{align}
\Delta_M\simeq&2\sqrt{\sum_{j,k} \left(\sum_{p,q}
|a^*_{k,p}a_{j,q}|\right)^2},\\
 \Delta_{\vec{C}}\simeq&2\sqrt{\sum_{j} \left(\sum_{p,q}
|a^*_{k,p}f_{j}|\right)^2}.
 \label{Eq:imaginaryshotnoiseDelta}
\end{align}
Using these quantities, we can rewrite Eq.~(\ref{Eq:imaginaryshotnoisenorm}) as
\begin{align}
\left|\delta\vec{\dot{\theta}}\right|
 \le&\frac{\left\|M_0^{-1}\right\|\Delta_{\vec{C}}+\left\|M_0^{-1}\right\|^2 |\vec{C}_0|\Delta_M}{\sqrt{N_m}}\\
 =&\frac{\Delta}{\sqrt{N_m}},
\end{align}
where we define $\Delta$ by
\begin{align}
\Delta=\left\|M_0^{-1}\right\|\Delta_{\vec{C}}+\left\|M_0^{-1}\right\|^2 |\vec{C}_0|\Delta_M.
\end{align}
By using the trace distance,
the previous study \cite{li2017efficient} expressed the degree of the error between the ideal wavefunction and the wavefunction obtained from the variational algorithms when we update from $\vec{\theta}(\tau)$ to $\vec{\theta}(\tau+\delta \tau)$ as follows:
\begin{align}
&D(\ket{\varphi^{(0)} (\vec{\theta}_0(\tau+\delta \tau))},\ket{\varphi (\vec{\theta}(\tau+\delta \tau))})\notag\\
&=\sqrt{\left[\delta\vec{\dot{\theta}}^TA(\tau)\delta\vec{\dot{\theta}}\right](\delta \tau)^2 +O((\delta \tau)^3)},
\end{align}
where $\ket{\varphi^{(0)} (\vec{\theta}_0(\tau+\delta \tau))}$ denotes the ideal wavefunction at time $\tau +\delta \tau$. Here, the matrix $A(\tau)$ is defined by
\begin{align}
&[A(\tau)]_{i,j}\notag\\
&= M_{i,j} - \frac{\partial\bra{\varphi (\vec{\theta}(\tau))}}{\partial \theta_j}\ket{\varphi (\vec{\theta}(\tau))}\bra{\varphi (\vec{\theta}(\tau))}\frac{\partial\ket{\varphi (\vec{\theta}(\tau))}}{\partial \theta_i}.
\end{align}
%Therefore, 
From these calculations,
the total error $D_I$ from $\tau =0$ to $\tau =T$ is given by
\begin{align}
D_I=&D(\ket{\varphi^{(0)} (\vec{\theta}_0(T))},\ket{\varphi (\vec{\theta}(T))})\\
\le &\sum_n D(\ket{\varphi^{(0)} (\vec{\theta}_0(n\delta \tau))},\ket{\varphi (\vec{\theta}(n \delta \tau))})\\
\lesssim& \sqrt{\|A\|_{{\rm max}}}\|\delta\vec{\dot{\theta}}\|_{{\rm max}}T\\
=&\sqrt{\|A\|_{{\rm max}}}\frac{\Delta_{{\rm max}} T}{\sqrt{N_m}},
\end{align}
where $\|\cdots\|_{{\rm max}}$ is the maximum value on the time interval $0\le \tau \le T$.
In order to obtain the expectation values within the accuracy $\varepsilon_s$, we need 
\begin{align}
\varepsilon_s \ge&\sqrt{\|A\|_{{\rm max}}}\frac{\Delta_{{\rm max}} T}{\sqrt{N_m}},
\end{align}
and we can obtain the condition on the required number of measurements as 
\begin{align}
N_m \ge \|A\|_{{\rm max}}\Delta_{{\rm max}}^2 T^2/\varepsilon_s^2.
\label{Eq:numberofmeasurements}
\end{align}
If this condition is satisfied by taking a large number of measurements,
%such that the number of measurements is sufficiently large, 
we obtain both
%both 
the KLD and the trace distance 
%are also obtained 
within the accuracy $\varepsilon_s$ except for a constant factor, according to Eq.~(\ref{Eq:Inequalities}).

\section{Comparison with contrastive divergence learning}
In this Appendix, we compare our scheme with contrastive divergence (CD) learning,
which is a standard technique to train the RBM~\cite{hinton2002training,carreira2005contrastive,bengio2009justifying}.
Although the CD learning is mainly used for the RBM, the learning algorithm can 
 be also applied to the fully visible BM~\cite{carreira2005contrastive}, such as our model in the main text.
Our calculation is restricted for the case of a few qubits because no actual NISQ device with relatively large qubits is currently available, as described in Section VI.

We apply the $k$-step CD learning (CD-$k$) for calculating the thermal average in Eq.~\eqref{Eq:derivative_of_KL}
to update the parameters in the BM.
To calculate the thermal average by using CD-$k$, we generate $D$ samples from the Markov chain with length $k$ and approximate the thermal average with the sample average.
An initial distribution of the Markov chain
is the uniform distribution of the training data set $\{\sigmavector^{(d)}\}_{d=1}^D$.
For simplicity, we assume that each data in the training data set appears once in the sampling from the initial distribution when generating the $D$ samples.
In the Markov process, we use the heat-bath algorithm to update the state, where the parameters are sequentially updated from $\sigma_1$ to $\sigma_N$.

\begin{figure}
  \includegraphics[width=0.5\textwidth]{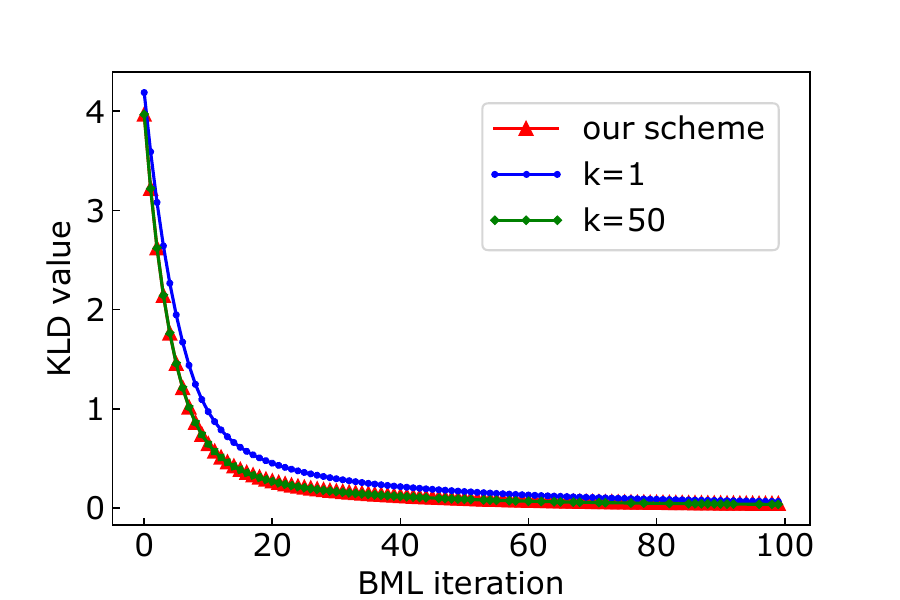}
   \caption{The KLD between the true distribution and the estimated distributions which are obtained from three different methods; our scheme, the CD-1 learning ($k=1$), and the CD-50 learning ($k=50$) at each iteration step. We take averages of the KLD over all group of training data at each step.}
  \label{fig:comparison with CD}
\end{figure}
Figure~\ref{fig:comparison with CD} shows the average of the KLD between the true distribution $P(\boldsymbol{\sigma}|\uvector^{*}_l)$ and the estimated probabilities (our scheme, the CD-1 learning, and the CD-50 learning)
at the $s$-th step over all the true parameter sets $\uvector^{*}_l$ $(l=1,2,\dots, L)$. 
We apply the common parameters of $N=4, L=30, D=10^4, N_{{\rm step}}=100$, and $\eta=0.1$.
These results show that while the values of the KLD in our scheme converge more rapidly than the ones in the CD-$1$ learning, the performance of our scheme is comparable with that of the CD-$50$. This is because both our scheme and the CD-50 learning achieve thermal average values with high accuracy compared with the CD-1 learning, and therefore both of the learning parameters are almost the same at each step $s$. The rapid convergence compared with the CD-$1$ learning probably depends on problems such as a data set and the number of qubits.
Therefore, in order to prove quantum advantage of our scheme over the
classical schemes, we need a careful benchmark of the performance using an actual NISQ device with relatively larger qubits.
We leave this for a future work.

%\begin{figure}
%\begin{align*}
%\Qcircuit @C=1em @R=.7em {
%&&&&&&&&&&&&\\
%\lstick{\ket{+}} &\qw&\gate{R_Y}&\ctrl{1}&\gate{R_Y}&\ctrl{1}
%&\qw&\qw&\qw&\qw&\qw&\gate{R_Y}&\qw\\
%\lstick{\ket{+}} &\qw&\gate{R_Y}&\ctrl{0}&\gate{R_Y}&\ctrl{0}
%&\gate{R_Y}&\ctrl{1}&\gate{R_Y}&\ctrl{1}&\qw&\gate{R_Y}&\qw\\
%\lstick{\ket{+}} &\qw&\gate{R_Y}&\ctrl{1}&\gate{R_Y}&\ctrl{1}
%&\gate{R_Y}&\ctrl{0}&\gate{R_Y}&\ctrl{0}&\qw&\gate{R_Y}&\qw\\
%\lstick{\ket{+}} &\qw&\gate{R_Y}&\ctrl{0}&\gate{R_Y}&\ctrl{0}
%&\qw&\qw&\qw&\qw&\qw&\gate{R_Y}&\qw\\
%&&&&&&&&&&&\\
%&&&&&&&&&&&\hspace{1em}\times N_{depth}
%\gategroup{1}{2}{6}{11}{0.7em}{--}
%}
%\nonumber
%\end{align*}
%\caption{\textcolor{green}{Temporary}}
%\label{fig:fail circuit}
%\end{figure}
\bibliographystyle{apsrev4-1}
\bibliography{shingu}

\begin{thebibliography}{75}%
\makeatletter
\providecommand \@ifxundefined [1]{%
 \@ifx{#1\undefined}
}%
\providecommand \@ifnum [1]{%
 \ifnum #1\expandafter \@firstoftwo
 \else \expandafter \@secondoftwo
 \fi
}%
\providecommand \@ifx [1]{%
 \ifx #1\expandafter \@firstoftwo
 \else \expandafter \@secondoftwo
 \fi
}%
\providecommand \natexlab [1]{#1}%
\providecommand \enquote  [1]{``#1''}%
\providecommand \bibnamefont  [1]{#1}%
\providecommand \bibfnamefont [1]{#1}%
\providecommand \citenamefont [1]{#1}%
\providecommand \href@noop [0]{\@secondoftwo}%
\providecommand \href [0]{\begingroup \@sanitize@url \@href}%
\providecommand \@href[1]{\@@startlink{#1}\@@href}%
\providecommand \@@href[1]{\endgroup#1\@@endlink}%
\providecommand \@sanitize@url [0]{\catcode `\\12\catcode `\$12\catcode
  `\&12\catcode `\#12\catcode `\^12\catcode `\_12\catcode `\%12\relax}%
\providecommand \@@startlink[1]{}%
\providecommand \@@endlink[0]{}%
\providecommand \url  [0]{\begingroup\@sanitize@url \@url }%
\providecommand \@url [1]{\endgroup\@href {#1}{\urlprefix }}%
\providecommand \urlprefix  [0]{URL }%
\providecommand \Eprint [0]{\href }%
\providecommand \doibase [0]{http://dx.doi.org/}%
\providecommand \selectlanguage [0]{\@gobble}%
\providecommand \bibinfo  [0]{\@secondoftwo}%
\providecommand \bibfield  [0]{\@secondoftwo}%
\providecommand \translation [1]{[#1]}%
\providecommand \BibitemOpen [0]{}%
\providecommand \bibitemStop [0]{}%
\providecommand \bibitemNoStop [0]{.\EOS\space}%
\providecommand \EOS [0]{\spacefactor3000\relax}%
\providecommand \BibitemShut  [1]{\csname bibitem#1\endcsname}%
\let\auto@bib@innerbib\@empty
%</preamble>
\bibitem [{\citenamefont {Fahlman}\ \emph {et~al.}(1983)\citenamefont
  {Fahlman}, \citenamefont {Hinton},\ and\ \citenamefont
  {Sejnowski}}]{fahlman1983massively}%
  \BibitemOpen
  \bibfield  {author} {\bibinfo {author} {\bibfnamefont {S.~E.}\ \bibnamefont
  {Fahlman}}, \bibinfo {author} {\bibfnamefont {G.~E.}\ \bibnamefont {Hinton}},
  \ and\ \bibinfo {author} {\bibfnamefont {T.~J.}\ \bibnamefont {Sejnowski}},\
  }in\ \href@noop {} {\emph {\bibinfo {booktitle} {National Conference on
  Artificial Intelligence, AAAI}}}\ (\bibinfo {year} {1983})\BibitemShut
  {NoStop}%
\bibitem [{\citenamefont {Ackley}\ \emph {et~al.}(1985)\citenamefont {Ackley},
  \citenamefont {Hinton},\ and\ \citenamefont
  {Sejnowski}}]{ackley1985learning}%
  \BibitemOpen
  \bibfield  {author} {\bibinfo {author} {\bibfnamefont {D.~H.}\ \bibnamefont
  {Ackley}}, \bibinfo {author} {\bibfnamefont {G.~E.}\ \bibnamefont {Hinton}},
  \ and\ \bibinfo {author} {\bibfnamefont {T.~J.}\ \bibnamefont {Sejnowski}},\
  }\href@noop {} {\bibfield  {journal} {\bibinfo  {journal} {Cognitive
  science}\ }\textbf {\bibinfo {volume} {9}},\ \bibinfo {pages} {147} (\bibinfo
  {year} {1985})}\BibitemShut {NoStop}%
\bibitem [{\citenamefont {Smolensky}(1986)}]{smolensky1986information}%
  \BibitemOpen
  \bibfield  {author} {\bibinfo {author} {\bibfnamefont {P.}~\bibnamefont
  {Smolensky}},\ }\href@noop {} {\emph {\bibinfo {title} {Information
  processing in dynamical systems: Foundations of harmony theory}}},\ \bibinfo
  {type} {Tech. Rep.}\ (\bibinfo  {institution} {Colorado Univ at Boulder Dept
  of Computer Science},\ \bibinfo {year} {1986})\BibitemShut {NoStop}%
\bibitem [{\citenamefont {Nair}\ and\ \citenamefont
  {Hinton}(2010)}]{nair2010rectified}%
  \BibitemOpen
  \bibfield  {author} {\bibinfo {author} {\bibfnamefont {V.}~\bibnamefont
  {Nair}}\ and\ \bibinfo {author} {\bibfnamefont {G.~E.}\ \bibnamefont
  {Hinton}},\ }in\ \href@noop {} {\emph {\bibinfo {booktitle} {Proceedings of
  the 27th international conference on machine learning (ICML-10)}}}\ (\bibinfo
  {year} {2010})\ pp.\ \bibinfo {pages} {807--814}\BibitemShut {NoStop}%
\bibitem [{\citenamefont {Hinton}(2012)}]{hinton2012practical}%
  \BibitemOpen
  \bibfield  {author} {\bibinfo {author} {\bibfnamefont {G.~E.}\ \bibnamefont
  {Hinton}},\ }in\ \href@noop {} {\emph {\bibinfo {booktitle} {Neural networks:
  Tricks of the trade}}}\ (\bibinfo  {publisher} {Springer},\ \bibinfo {year}
  {2012})\ pp.\ \bibinfo {pages} {599--619}\BibitemShut {NoStop}%
\bibitem [{\citenamefont {Le~Roux}\ and\ \citenamefont
  {Bengio}(2008)}]{le2008representational}%
  \BibitemOpen
  \bibfield  {author} {\bibinfo {author} {\bibfnamefont {N.}~\bibnamefont
  {Le~Roux}}\ and\ \bibinfo {author} {\bibfnamefont {Y.}~\bibnamefont
  {Bengio}},\ }\href@noop {} {\bibfield  {journal} {\bibinfo  {journal} {Neural
  computation}\ }\textbf {\bibinfo {volume} {20}},\ \bibinfo {pages} {1631}
  (\bibinfo {year} {2008})}\BibitemShut {NoStop}%
\bibitem [{\citenamefont {Hinton}\ and\ \citenamefont
  {Salakhutdinov}(2006)}]{hinton2006reducing}%
  \BibitemOpen
  \bibfield  {author} {\bibinfo {author} {\bibfnamefont {G.~E.}\ \bibnamefont
  {Hinton}}\ and\ \bibinfo {author} {\bibfnamefont {R.~R.}\ \bibnamefont
  {Salakhutdinov}},\ }\href@noop {} {\bibfield  {journal} {\bibinfo  {journal}
  {science}\ }\textbf {\bibinfo {volume} {313}},\ \bibinfo {pages} {504}
  (\bibinfo {year} {2006})}\BibitemShut {NoStop}%
\bibitem [{\citenamefont {Salakhutdinov}\ \emph {et~al.}(2007)\citenamefont
  {Salakhutdinov}, \citenamefont {Mnih},\ and\ \citenamefont
  {Hinton}}]{salakhutdinov2007restricted}%
  \BibitemOpen
  \bibfield  {author} {\bibinfo {author} {\bibfnamefont {R.}~\bibnamefont
  {Salakhutdinov}}, \bibinfo {author} {\bibfnamefont {A.}~\bibnamefont {Mnih}},
  \ and\ \bibinfo {author} {\bibfnamefont {G.}~\bibnamefont {Hinton}},\ }in\
  \href@noop {} {\emph {\bibinfo {booktitle} {Proceedings of the 24th
  international conference on Machine learning}}}\ (\bibinfo {year} {2007})\
  pp.\ \bibinfo {pages} {791--798}\BibitemShut {NoStop}%
\bibitem [{\citenamefont {Larochelle}\ and\ \citenamefont
  {Bengio}(2008)}]{larochelle2008classification}%
  \BibitemOpen
  \bibfield  {author} {\bibinfo {author} {\bibfnamefont {H.}~\bibnamefont
  {Larochelle}}\ and\ \bibinfo {author} {\bibfnamefont {Y.}~\bibnamefont
  {Bengio}},\ }in\ \href@noop {} {\emph {\bibinfo {booktitle} {Proceedings of
  the 25th international conference on Machine learning}}}\ (\bibinfo {year}
  {2008})\ pp.\ \bibinfo {pages} {536--543}\BibitemShut {NoStop}%
\bibitem [{\citenamefont {Melko}\ \emph {et~al.}(2019)\citenamefont {Melko},
  \citenamefont {Carleo}, \citenamefont {Carrasquilla},\ and\ \citenamefont
  {Cirac}}]{melko2019restricted}%
  \BibitemOpen
  \bibfield  {author} {\bibinfo {author} {\bibfnamefont {R.~G.}\ \bibnamefont
  {Melko}}, \bibinfo {author} {\bibfnamefont {G.}~\bibnamefont {Carleo}},
  \bibinfo {author} {\bibfnamefont {J.}~\bibnamefont {Carrasquilla}}, \ and\
  \bibinfo {author} {\bibfnamefont {J.~I.}\ \bibnamefont {Cirac}},\ }\href@noop
  {} {\bibfield  {journal} {\bibinfo  {journal} {Nature Physics}\ }\textbf
  {\bibinfo {volume} {15}},\ \bibinfo {pages} {887} (\bibinfo {year}
  {2019})}\BibitemShut {NoStop}%
\bibitem [{\citenamefont {Jia}\ \emph {et~al.}(2019)\citenamefont {Jia},
  \citenamefont {Yi}, \citenamefont {Zhai}, \citenamefont {Wu}, \citenamefont
  {Guo},\ and\ \citenamefont {Guo}}]{jia2019quantum}%
  \BibitemOpen
  \bibfield  {author} {\bibinfo {author} {\bibfnamefont {Z.-A.}\ \bibnamefont
  {Jia}}, \bibinfo {author} {\bibfnamefont {B.}~\bibnamefont {Yi}}, \bibinfo
  {author} {\bibfnamefont {R.}~\bibnamefont {Zhai}}, \bibinfo {author}
  {\bibfnamefont {Y.-C.}\ \bibnamefont {Wu}}, \bibinfo {author} {\bibfnamefont
  {G.-C.}\ \bibnamefont {Guo}}, \ and\ \bibinfo {author} {\bibfnamefont
  {G.-P.}\ \bibnamefont {Guo}},\ }\href@noop {} {\bibfield  {journal} {\bibinfo
   {journal} {Advanced Quantum Technologies}\ }\textbf {\bibinfo {volume}
  {2}},\ \bibinfo {pages} {1800077} (\bibinfo {year} {2019})}\BibitemShut
  {NoStop}%
\bibitem [{\citenamefont {Hsieh}\ \emph {et~al.}(2019)\citenamefont {Hsieh},
  \citenamefont {Sun}, \citenamefont {Zhang},\ and\ \citenamefont
  {Lee}}]{hsieh2019unitary}%
  \BibitemOpen
  \bibfield  {author} {\bibinfo {author} {\bibfnamefont {C.-y.}\ \bibnamefont
  {Hsieh}}, \bibinfo {author} {\bibfnamefont {Q.}~\bibnamefont {Sun}}, \bibinfo
  {author} {\bibfnamefont {S.}~\bibnamefont {Zhang}}, \ and\ \bibinfo {author}
  {\bibfnamefont {C.~K.}\ \bibnamefont {Lee}},\ }\href@noop {} {\bibfield
  {journal} {\bibinfo  {journal} {arXiv preprint arXiv:1912.02988}\ } (\bibinfo
  {year} {2019})}\BibitemShut {NoStop}%
\bibitem [{\citenamefont {Hinton}\ and\ \citenamefont
  {Salakhutdinov}(2009)}]{hinton2009replicated}%
  \BibitemOpen
  \bibfield  {author} {\bibinfo {author} {\bibfnamefont {G.~E.}\ \bibnamefont
  {Hinton}}\ and\ \bibinfo {author} {\bibfnamefont {R.~R.}\ \bibnamefont
  {Salakhutdinov}},\ }in\ \href@noop {} {\emph {\bibinfo {booktitle} {Advances
  in neural information processing systems}}}\ (\bibinfo {year} {2009})\ pp.\
  \bibinfo {pages} {1607--1614}\BibitemShut {NoStop}%
\bibitem [{\citenamefont {Coates}\ \emph {et~al.}(2011)\citenamefont {Coates},
  \citenamefont {Ng},\ and\ \citenamefont {Lee}}]{coates2011analysis}%
  \BibitemOpen
  \bibfield  {author} {\bibinfo {author} {\bibfnamefont {A.}~\bibnamefont
  {Coates}}, \bibinfo {author} {\bibfnamefont {A.}~\bibnamefont {Ng}}, \ and\
  \bibinfo {author} {\bibfnamefont {H.}~\bibnamefont {Lee}},\ }in\ \href@noop
  {} {\emph {\bibinfo {booktitle} {Proceedings of the fourteenth international
  conference on artificial intelligence and statistics}}}\ (\bibinfo {year}
  {2011})\ pp.\ \bibinfo {pages} {215--223}\BibitemShut {NoStop}%
\bibitem [{\citenamefont {Dahl}\ \emph {et~al.}(2011)\citenamefont {Dahl},
  \citenamefont {Yu}, \citenamefont {Deng},\ and\ \citenamefont
  {Acero}}]{dahl2011context}%
  \BibitemOpen
  \bibfield  {author} {\bibinfo {author} {\bibfnamefont {G.~E.}\ \bibnamefont
  {Dahl}}, \bibinfo {author} {\bibfnamefont {D.}~\bibnamefont {Yu}}, \bibinfo
  {author} {\bibfnamefont {L.}~\bibnamefont {Deng}}, \ and\ \bibinfo {author}
  {\bibfnamefont {A.}~\bibnamefont {Acero}},\ }\href@noop {} {\bibfield
  {journal} {\bibinfo  {journal} {IEEE Transactions on audio, speech, and
  language processing}\ }\textbf {\bibinfo {volume} {20}},\ \bibinfo {pages}
  {30} (\bibinfo {year} {2011})}\BibitemShut {NoStop}%
\bibitem [{\citenamefont {Hinton}\ \emph {et~al.}(2006)\citenamefont {Hinton},
  \citenamefont {Osindero},\ and\ \citenamefont {Teh}}]{hinton2006fast}%
  \BibitemOpen
  \bibfield  {author} {\bibinfo {author} {\bibfnamefont {G.~E.}\ \bibnamefont
  {Hinton}}, \bibinfo {author} {\bibfnamefont {S.}~\bibnamefont {Osindero}}, \
  and\ \bibinfo {author} {\bibfnamefont {Y.-W.}\ \bibnamefont {Teh}},\
  }\href@noop {} {\bibfield  {journal} {\bibinfo  {journal} {Neural
  computation}\ }\textbf {\bibinfo {volume} {18}},\ \bibinfo {pages} {1527}
  (\bibinfo {year} {2006})}\BibitemShut {NoStop}%
\bibitem [{\citenamefont {Salakhutdinov}\ and\ \citenamefont
  {Hinton}(2009)}]{salakhutdinov2009deep}%
  \BibitemOpen
  \bibfield  {author} {\bibinfo {author} {\bibfnamefont {R.}~\bibnamefont
  {Salakhutdinov}}\ and\ \bibinfo {author} {\bibfnamefont {G.}~\bibnamefont
  {Hinton}},\ }in\ \href@noop {} {\emph {\bibinfo {booktitle} {Artificial
  intelligence and statistics}}}\ (\bibinfo {organization} {PMLR},\ \bibinfo
  {year} {2009})\ pp.\ \bibinfo {pages} {448--455}\BibitemShut {NoStop}%
\bibitem [{\citenamefont {Salakhutdinov}\ and\ \citenamefont
  {Hinton}(2012)}]{salakhutdinov2012efficient}%
  \BibitemOpen
  \bibfield  {author} {\bibinfo {author} {\bibfnamefont {R.}~\bibnamefont
  {Salakhutdinov}}\ and\ \bibinfo {author} {\bibfnamefont {G.}~\bibnamefont
  {Hinton}},\ }\href@noop {} {\bibfield  {journal} {\bibinfo  {journal} {Neural
  computation}\ }\textbf {\bibinfo {volume} {24}},\ \bibinfo {pages} {1967}
  (\bibinfo {year} {2012})}\BibitemShut {NoStop}%
\bibitem [{\citenamefont {Hinton}(2002)}]{hinton2002training}%
  \BibitemOpen
  \bibfield  {author} {\bibinfo {author} {\bibfnamefont {G.~E.}\ \bibnamefont
  {Hinton}},\ }\href@noop {} {\bibfield  {journal} {\bibinfo  {journal} {Neural
  computation}\ }\textbf {\bibinfo {volume} {14}},\ \bibinfo {pages} {1771}
  (\bibinfo {year} {2002})}\BibitemShut {NoStop}%
\bibitem [{\citenamefont {Carreira-Perpinan}\ and\ \citenamefont
  {Hinton}(2005)}]{carreira2005contrastive}%
  \BibitemOpen
  \bibfield  {author} {\bibinfo {author} {\bibfnamefont {M.~A.}\ \bibnamefont
  {Carreira-Perpinan}}\ and\ \bibinfo {author} {\bibfnamefont {G.~E.}\
  \bibnamefont {Hinton}},\ }in\ \href@noop {} {\emph {\bibinfo {booktitle}
  {Aistats}}},\ Vol.~\bibinfo {volume} {10}\ (\bibinfo {organization}
  {Citeseer},\ \bibinfo {year} {2005})\ pp.\ \bibinfo {pages}
  {33--40}\BibitemShut {NoStop}%
\bibitem [{\citenamefont {Bengio}\ and\ \citenamefont
  {Delalleau}(2009)}]{bengio2009justifying}%
  \BibitemOpen
  \bibfield  {author} {\bibinfo {author} {\bibfnamefont {Y.}~\bibnamefont
  {Bengio}}\ and\ \bibinfo {author} {\bibfnamefont {O.}~\bibnamefont
  {Delalleau}},\ }\href@noop {} {\bibfield  {journal} {\bibinfo  {journal}
  {Neural computation}\ }\textbf {\bibinfo {volume} {21}},\ \bibinfo {pages}
  {1601} (\bibinfo {year} {2009})}\BibitemShut {NoStop}%
\bibitem [{\citenamefont {Gabri{\'e}}\ \emph {et~al.}(2015)\citenamefont
  {Gabri{\'e}}, \citenamefont {Tramel},\ and\ \citenamefont
  {Krzakala}}]{gabrie2015training}%
  \BibitemOpen
  \bibfield  {author} {\bibinfo {author} {\bibfnamefont {M.}~\bibnamefont
  {Gabri{\'e}}}, \bibinfo {author} {\bibfnamefont {E.~W.}\ \bibnamefont
  {Tramel}}, \ and\ \bibinfo {author} {\bibfnamefont {F.}~\bibnamefont
  {Krzakala}},\ }in\ \href@noop {} {\emph {\bibinfo {booktitle} {Advances in
  neural information processing systems}}}\ (\bibinfo {year} {2015})\ pp.\
  \bibinfo {pages} {640--648}\BibitemShut {NoStop}%
\bibitem [{\citenamefont {Marlin}\ \emph {et~al.}(2010)\citenamefont {Marlin},
  \citenamefont {Swersky}, \citenamefont {Chen},\ and\ \citenamefont
  {Freitas}}]{marlin2010inductive}%
  \BibitemOpen
  \bibfield  {author} {\bibinfo {author} {\bibfnamefont {B.}~\bibnamefont
  {Marlin}}, \bibinfo {author} {\bibfnamefont {K.}~\bibnamefont {Swersky}},
  \bibinfo {author} {\bibfnamefont {B.}~\bibnamefont {Chen}}, \ and\ \bibinfo
  {author} {\bibfnamefont {N.}~\bibnamefont {Freitas}},\ }in\ \href@noop {}
  {\emph {\bibinfo {booktitle} {Proceedings of the Thirteenth International
  Conference on Artificial Intelligence and Statistics}}}\ (\bibinfo {year}
  {2010})\ pp.\ \bibinfo {pages} {509--516}\BibitemShut {NoStop}%
\bibitem [{\citenamefont {Yasuda}\ \emph {et~al.}(2012)\citenamefont {Yasuda},
  \citenamefont {Kataoka}, \citenamefont {Waizumi},\ and\ \citenamefont
  {Tanaka}}]{yasuda2012composite}%
  \BibitemOpen
  \bibfield  {author} {\bibinfo {author} {\bibfnamefont {M.}~\bibnamefont
  {Yasuda}}, \bibinfo {author} {\bibfnamefont {S.}~\bibnamefont {Kataoka}},
  \bibinfo {author} {\bibfnamefont {Y.}~\bibnamefont {Waizumi}}, \ and\
  \bibinfo {author} {\bibfnamefont {K.}~\bibnamefont {Tanaka}},\ }in\
  \href@noop {} {\emph {\bibinfo {booktitle} {Proceedings of the 21st
  International Conference on Pattern Recognition (ICPR2012)}}}\ (\bibinfo
  {organization} {IEEE},\ \bibinfo {year} {2012})\ pp.\ \bibinfo {pages}
  {2234--2237}\BibitemShut {NoStop}%
\bibitem [{\citenamefont {Yasuda}\ \emph {et~al.}(2011)\citenamefont {Yasuda},
  \citenamefont {Sakurai},\ and\ \citenamefont {Tanaka}}]{yasuda2011learning}%
  \BibitemOpen
  \bibfield  {author} {\bibinfo {author} {\bibfnamefont {M.}~\bibnamefont
  {Yasuda}}, \bibinfo {author} {\bibfnamefont {T.}~\bibnamefont {Sakurai}}, \
  and\ \bibinfo {author} {\bibfnamefont {K.}~\bibnamefont {Tanaka}},\
  }\href@noop {} {\bibfield  {journal} {\bibinfo  {journal} {Nonlinear Theory
  and Its Applications, IEICE}\ }\textbf {\bibinfo {volume} {2}},\ \bibinfo
  {pages} {153} (\bibinfo {year} {2011})}\BibitemShut {NoStop}%
\bibitem [{\citenamefont {Rebentrost}\ \emph {et~al.}(2014)\citenamefont
  {Rebentrost}, \citenamefont {Mohseni},\ and\ \citenamefont
  {Lloyd}}]{rebentrost2014quantum}%
  \BibitemOpen
  \bibfield  {author} {\bibinfo {author} {\bibfnamefont {P.}~\bibnamefont
  {Rebentrost}}, \bibinfo {author} {\bibfnamefont {M.}~\bibnamefont {Mohseni}},
  \ and\ \bibinfo {author} {\bibfnamefont {S.}~\bibnamefont {Lloyd}},\
  }\href@noop {} {\bibfield  {journal} {\bibinfo  {journal} {Physical review
  letters}\ }\textbf {\bibinfo {volume} {113}},\ \bibinfo {pages} {130503}
  (\bibinfo {year} {2014})}\BibitemShut {NoStop}%
\bibitem [{\citenamefont {Schuld}\ \emph {et~al.}(2016)\citenamefont {Schuld},
  \citenamefont {Sinayskiy},\ and\ \citenamefont
  {Petruccione}}]{schuld2016prediction}%
  \BibitemOpen
  \bibfield  {author} {\bibinfo {author} {\bibfnamefont {M.}~\bibnamefont
  {Schuld}}, \bibinfo {author} {\bibfnamefont {I.}~\bibnamefont {Sinayskiy}}, \
  and\ \bibinfo {author} {\bibfnamefont {F.}~\bibnamefont {Petruccione}},\
  }\href@noop {} {\bibfield  {journal} {\bibinfo  {journal} {Physical Review
  A}\ }\textbf {\bibinfo {volume} {94}},\ \bibinfo {pages} {022342} (\bibinfo
  {year} {2016})}\BibitemShut {NoStop}%
\bibitem [{\citenamefont {Wiebe}\ \emph {et~al.}(2012)\citenamefont {Wiebe},
  \citenamefont {Braun},\ and\ \citenamefont {Lloyd}}]{wiebe2012quantum}%
  \BibitemOpen
  \bibfield  {author} {\bibinfo {author} {\bibfnamefont {N.}~\bibnamefont
  {Wiebe}}, \bibinfo {author} {\bibfnamefont {D.}~\bibnamefont {Braun}}, \ and\
  \bibinfo {author} {\bibfnamefont {S.}~\bibnamefont {Lloyd}},\ }\href@noop {}
  {\bibfield  {journal} {\bibinfo  {journal} {Physical review letters}\
  }\textbf {\bibinfo {volume} {109}},\ \bibinfo {pages} {050505} (\bibinfo
  {year} {2012})}\BibitemShut {NoStop}%
\bibitem [{\citenamefont {Lloyd}\ \emph {et~al.}(2014)\citenamefont {Lloyd},
  \citenamefont {Mohseni},\ and\ \citenamefont
  {Rebentrost}}]{lloyd2014quantum}%
  \BibitemOpen
  \bibfield  {author} {\bibinfo {author} {\bibfnamefont {S.}~\bibnamefont
  {Lloyd}}, \bibinfo {author} {\bibfnamefont {M.}~\bibnamefont {Mohseni}}, \
  and\ \bibinfo {author} {\bibfnamefont {P.}~\bibnamefont {Rebentrost}},\
  }\href@noop {} {\bibfield  {journal} {\bibinfo  {journal} {Nature Physics}\
  }\textbf {\bibinfo {volume} {10}},\ \bibinfo {pages} {631} (\bibinfo {year}
  {2014})}\BibitemShut {NoStop}%
\bibitem [{\citenamefont {Preskill}(2018)}]{preskill2018quantum}%
  \BibitemOpen
  \bibfield  {author} {\bibinfo {author} {\bibfnamefont {J.}~\bibnamefont
  {Preskill}},\ }\href@noop {} {\bibfield  {journal} {\bibinfo  {journal}
  {Quantum}\ }\textbf {\bibinfo {volume} {2}},\ \bibinfo {pages} {79} (\bibinfo
  {year} {2018})}\BibitemShut {NoStop}%
\bibitem [{\citenamefont {Endo}\ \emph {et~al.}(2021)\citenamefont {Endo},
  \citenamefont {Cai}, \citenamefont {Benjamin},\ and\ \citenamefont
  {Yuan}}]{doi:10.7566/JPSJ.90.032001}%
  \BibitemOpen
  \bibfield  {author} {\bibinfo {author} {\bibfnamefont {S.}~\bibnamefont
  {Endo}}, \bibinfo {author} {\bibfnamefont {Z.}~\bibnamefont {Cai}}, \bibinfo
  {author} {\bibfnamefont {S.~C.}\ \bibnamefont {Benjamin}}, \ and\ \bibinfo
  {author} {\bibfnamefont {X.}~\bibnamefont {Yuan}},\ }\href {\doibase
  10.7566/JPSJ.90.032001} {\bibfield  {journal} {\bibinfo  {journal} {Journal
  of the Physical Society of Japan}\ }\textbf {\bibinfo {volume} {90}},\
  \bibinfo {pages} {032001} (\bibinfo {year} {2021})}\BibitemShut {NoStop}%
\bibitem [{\citenamefont {Cerezo}\ \emph {et~al.}(2020)\citenamefont {Cerezo},
  \citenamefont {Arrasmith}, \citenamefont {Babbush}, \citenamefont {Benjamin},
  \citenamefont {Endo}, \citenamefont {Fujii}, \citenamefont {McClean},
  \citenamefont {Mitarai}, \citenamefont {Yuan}, \citenamefont {Cincio} \emph
  {et~al.}}]{cerezo2020variational2}%
  \BibitemOpen
  \bibfield  {author} {\bibinfo {author} {\bibfnamefont {M.}~\bibnamefont
  {Cerezo}}, \bibinfo {author} {\bibfnamefont {A.}~\bibnamefont {Arrasmith}},
  \bibinfo {author} {\bibfnamefont {R.}~\bibnamefont {Babbush}}, \bibinfo
  {author} {\bibfnamefont {S.~C.}\ \bibnamefont {Benjamin}}, \bibinfo {author}
  {\bibfnamefont {S.}~\bibnamefont {Endo}}, \bibinfo {author} {\bibfnamefont
  {K.}~\bibnamefont {Fujii}}, \bibinfo {author} {\bibfnamefont {J.~R.}\
  \bibnamefont {McClean}}, \bibinfo {author} {\bibfnamefont {K.}~\bibnamefont
  {Mitarai}}, \bibinfo {author} {\bibfnamefont {X.}~\bibnamefont {Yuan}},
  \bibinfo {author} {\bibfnamefont {L.}~\bibnamefont {Cincio}},  \emph
  {et~al.},\ }\href@noop {} {\bibfield  {journal} {\bibinfo  {journal} {arXiv
  preprint arXiv:2012.09265}\ } (\bibinfo {year} {2020})}\BibitemShut {NoStop}%
\bibitem [{\citenamefont {Peruzzo}\ \emph {et~al.}(2014)\citenamefont
  {Peruzzo}, \citenamefont {McClean}, \citenamefont {Shadbolt}, \citenamefont
  {Yung}, \citenamefont {Zhou}, \citenamefont {Love}, \citenamefont
  {Aspuru-Guzik},\ and\ \citenamefont {O’brien}}]{peruzzo2014variational}%
  \BibitemOpen
  \bibfield  {author} {\bibinfo {author} {\bibfnamefont {A.}~\bibnamefont
  {Peruzzo}}, \bibinfo {author} {\bibfnamefont {J.}~\bibnamefont {McClean}},
  \bibinfo {author} {\bibfnamefont {P.}~\bibnamefont {Shadbolt}}, \bibinfo
  {author} {\bibfnamefont {M.-H.}\ \bibnamefont {Yung}}, \bibinfo {author}
  {\bibfnamefont {X.-Q.}\ \bibnamefont {Zhou}}, \bibinfo {author}
  {\bibfnamefont {P.~J.}\ \bibnamefont {Love}}, \bibinfo {author}
  {\bibfnamefont {A.}~\bibnamefont {Aspuru-Guzik}}, \ and\ \bibinfo {author}
  {\bibfnamefont {J.~L.}\ \bibnamefont {O’brien}},\ }\href@noop {} {\bibfield
   {journal} {\bibinfo  {journal} {Nature communications}\ }\textbf {\bibinfo
  {volume} {5}},\ \bibinfo {pages} {4213} (\bibinfo {year} {2014})}\BibitemShut
  {NoStop}%
\bibitem [{\citenamefont {Li}\ and\ \citenamefont
  {Benjamin}(2017)}]{li2017efficient}%
  \BibitemOpen
  \bibfield  {author} {\bibinfo {author} {\bibfnamefont {Y.}~\bibnamefont
  {Li}}\ and\ \bibinfo {author} {\bibfnamefont {S.~C.}\ \bibnamefont
  {Benjamin}},\ }\href@noop {} {\bibfield  {journal} {\bibinfo  {journal}
  {Physical Review X}\ }\textbf {\bibinfo {volume} {7}},\ \bibinfo {pages}
  {021050} (\bibinfo {year} {2017})}\BibitemShut {NoStop}%
\bibitem [{\citenamefont {McClean}\ \emph {et~al.}(2016)\citenamefont
  {McClean}, \citenamefont {Romero}, \citenamefont {Babbush},\ and\
  \citenamefont {Aspuru-Guzik}}]{mcclean2016theory}%
  \BibitemOpen
  \bibfield  {author} {\bibinfo {author} {\bibfnamefont {J.~R.}\ \bibnamefont
  {McClean}}, \bibinfo {author} {\bibfnamefont {J.}~\bibnamefont {Romero}},
  \bibinfo {author} {\bibfnamefont {R.}~\bibnamefont {Babbush}}, \ and\
  \bibinfo {author} {\bibfnamefont {A.}~\bibnamefont {Aspuru-Guzik}},\
  }\href@noop {} {\bibfield  {journal} {\bibinfo  {journal} {New Journal of
  Physics}\ }\textbf {\bibinfo {volume} {18}},\ \bibinfo {pages} {023023}
  (\bibinfo {year} {2016})}\BibitemShut {NoStop}%
\bibitem [{\citenamefont {Yuan}\ \emph {et~al.}(2019)\citenamefont {Yuan},
  \citenamefont {Endo}, \citenamefont {Zhao}, \citenamefont {Li},\ and\
  \citenamefont {Benjamin}}]{yuan2019theory}%
  \BibitemOpen
  \bibfield  {author} {\bibinfo {author} {\bibfnamefont {X.}~\bibnamefont
  {Yuan}}, \bibinfo {author} {\bibfnamefont {S.}~\bibnamefont {Endo}}, \bibinfo
  {author} {\bibfnamefont {Q.}~\bibnamefont {Zhao}}, \bibinfo {author}
  {\bibfnamefont {Y.}~\bibnamefont {Li}}, \ and\ \bibinfo {author}
  {\bibfnamefont {S.~C.}\ \bibnamefont {Benjamin}},\ }\href@noop {} {\bibfield
  {journal} {\bibinfo  {journal} {Quantum}\ }\textbf {\bibinfo {volume} {3}},\
  \bibinfo {pages} {191} (\bibinfo {year} {2019})}\BibitemShut {NoStop}%
\bibitem [{\citenamefont {Endo}\ \emph
  {et~al.}(2020{\natexlab{a}})\citenamefont {Endo}, \citenamefont {Sun},
  \citenamefont {Li}, \citenamefont {Benjamin},\ and\ \citenamefont
  {Yuan}}]{endo2018variational}%
  \BibitemOpen
  \bibfield  {author} {\bibinfo {author} {\bibfnamefont {S.}~\bibnamefont
  {Endo}}, \bibinfo {author} {\bibfnamefont {J.}~\bibnamefont {Sun}}, \bibinfo
  {author} {\bibfnamefont {Y.}~\bibnamefont {Li}}, \bibinfo {author}
  {\bibfnamefont {S.~C.}\ \bibnamefont {Benjamin}}, \ and\ \bibinfo {author}
  {\bibfnamefont {X.}~\bibnamefont {Yuan}},\ }\href {\doibase
  10.1103/PhysRevLett.125.010501} {\bibfield  {journal} {\bibinfo  {journal}
  {Phys. Rev. Lett.}\ }\textbf {\bibinfo {volume} {125}},\ \bibinfo {pages}
  {010501} (\bibinfo {year} {2020}{\natexlab{a}})}\BibitemShut {NoStop}%
\bibitem [{\citenamefont {Mitarai}\ \emph {et~al.}(2018)\citenamefont
  {Mitarai}, \citenamefont {Negoro}, \citenamefont {Kitagawa},\ and\
  \citenamefont {Fujii}}]{mitarai2018quantum}%
  \BibitemOpen
  \bibfield  {author} {\bibinfo {author} {\bibfnamefont {K.}~\bibnamefont
  {Mitarai}}, \bibinfo {author} {\bibfnamefont {M.}~\bibnamefont {Negoro}},
  \bibinfo {author} {\bibfnamefont {M.}~\bibnamefont {Kitagawa}}, \ and\
  \bibinfo {author} {\bibfnamefont {K.}~\bibnamefont {Fujii}},\ }\href@noop {}
  {\bibfield  {journal} {\bibinfo  {journal} {Physical Review A}\ }\textbf
  {\bibinfo {volume} {98}},\ \bibinfo {pages} {032309} (\bibinfo {year}
  {2018})}\BibitemShut {NoStop}%
\bibitem [{\citenamefont {Benedetti}\ \emph {et~al.}(2019)\citenamefont
  {Benedetti}, \citenamefont {Garcia-Pintos}, \citenamefont {Perdomo},
  \citenamefont {Leyton-Ortega}, \citenamefont {Nam},\ and\ \citenamefont
  {Perdomo-Ortiz}}]{benedetti2019generative}%
  \BibitemOpen
  \bibfield  {author} {\bibinfo {author} {\bibfnamefont {M.}~\bibnamefont
  {Benedetti}}, \bibinfo {author} {\bibfnamefont {D.}~\bibnamefont
  {Garcia-Pintos}}, \bibinfo {author} {\bibfnamefont {O.}~\bibnamefont
  {Perdomo}}, \bibinfo {author} {\bibfnamefont {V.}~\bibnamefont
  {Leyton-Ortega}}, \bibinfo {author} {\bibfnamefont {Y.}~\bibnamefont {Nam}},
  \ and\ \bibinfo {author} {\bibfnamefont {A.}~\bibnamefont {Perdomo-Ortiz}},\
  }\href@noop {} {\bibfield  {journal} {\bibinfo  {journal} {npj Quantum
  Information}\ }\textbf {\bibinfo {volume} {5}},\ \bibinfo {pages} {1}
  (\bibinfo {year} {2019})}\BibitemShut {NoStop}%
\bibitem [{\citenamefont {Amin}\ \emph {et~al.}(2018)\citenamefont {Amin},
  \citenamefont {Andriyash}, \citenamefont {Rolfe}, \citenamefont
  {Kulchytskyy},\ and\ \citenamefont {Melko}}]{amin2018quantum}%
  \BibitemOpen
  \bibfield  {author} {\bibinfo {author} {\bibfnamefont {M.~H.}\ \bibnamefont
  {Amin}}, \bibinfo {author} {\bibfnamefont {E.}~\bibnamefont {Andriyash}},
  \bibinfo {author} {\bibfnamefont {J.}~\bibnamefont {Rolfe}}, \bibinfo
  {author} {\bibfnamefont {B.}~\bibnamefont {Kulchytskyy}}, \ and\ \bibinfo
  {author} {\bibfnamefont {R.}~\bibnamefont {Melko}},\ }\href@noop {}
  {\bibfield  {journal} {\bibinfo  {journal} {Physical Review X}\ }\textbf
  {\bibinfo {volume} {8}},\ \bibinfo {pages} {021050} (\bibinfo {year}
  {2018})}\BibitemShut {NoStop}%
\bibitem [{\citenamefont {McArdle}\ \emph
  {et~al.}(2019{\natexlab{a}})\citenamefont {McArdle}, \citenamefont {Jones},
  \citenamefont {Endo}, \citenamefont {Li}, \citenamefont {Benjamin},\ and\
  \citenamefont {Yuan}}]{mcardle2019variational}%
  \BibitemOpen
  \bibfield  {author} {\bibinfo {author} {\bibfnamefont {S.}~\bibnamefont
  {McArdle}}, \bibinfo {author} {\bibfnamefont {T.}~\bibnamefont {Jones}},
  \bibinfo {author} {\bibfnamefont {S.}~\bibnamefont {Endo}}, \bibinfo {author}
  {\bibfnamefont {Y.}~\bibnamefont {Li}}, \bibinfo {author} {\bibfnamefont
  {S.~C.}\ \bibnamefont {Benjamin}}, \ and\ \bibinfo {author} {\bibfnamefont
  {X.}~\bibnamefont {Yuan}},\ }\href@noop {} {\bibfield  {journal} {\bibinfo
  {journal} {npj Quantum Information}\ }\textbf {\bibinfo {volume} {5}},\
  \bibinfo {pages} {1} (\bibinfo {year} {2019}{\natexlab{a}})}\BibitemShut
  {NoStop}%
\bibitem [{\citenamefont {McLachlan}(1964)}]{mclachlan1964variational}%
  \BibitemOpen
  \bibfield  {author} {\bibinfo {author} {\bibfnamefont {A.}~\bibnamefont
  {McLachlan}},\ }\href@noop {} {\bibfield  {journal} {\bibinfo  {journal}
  {Molecular Physics}\ }\textbf {\bibinfo {volume} {8}},\ \bibinfo {pages} {39}
  (\bibinfo {year} {1964})}\BibitemShut {NoStop}%
\bibitem [{\citenamefont {Broeckhove}\ \emph {et~al.}(1988)\citenamefont
  {Broeckhove}, \citenamefont {Lathouwers}, \citenamefont {Kesteloot},\ and\
  \citenamefont {Van~Leuven}}]{broeckhove1988equivalence}%
  \BibitemOpen
  \bibfield  {author} {\bibinfo {author} {\bibfnamefont {J.}~\bibnamefont
  {Broeckhove}}, \bibinfo {author} {\bibfnamefont {L.}~\bibnamefont
  {Lathouwers}}, \bibinfo {author} {\bibfnamefont {E.}~\bibnamefont
  {Kesteloot}}, \ and\ \bibinfo {author} {\bibfnamefont {P.}~\bibnamefont
  {Van~Leuven}},\ }\href@noop {} {\bibfield  {journal} {\bibinfo  {journal}
  {Chemical physics letters}\ }\textbf {\bibinfo {volume} {149}},\ \bibinfo
  {pages} {547} (\bibinfo {year} {1988})}\BibitemShut {NoStop}%
\bibitem [{\citenamefont {Nielsen}\ and\ \citenamefont
  {Chuang}(2002)}]{nielsen2002quantum}%
  \BibitemOpen
  \bibfield  {author} {\bibinfo {author} {\bibfnamefont {M.~A.}\ \bibnamefont
  {Nielsen}}\ and\ \bibinfo {author} {\bibfnamefont {I.}~\bibnamefont
  {Chuang}},\ }\href@noop {} {\enquote {\bibinfo {title} {Quantum computation
  and quantum information},}\ } (\bibinfo {year} {2002})\BibitemShut {NoStop}%
\bibitem [{\citenamefont {Endo}(2019)}]{endo2019hybrid}%
  \BibitemOpen
  \bibfield  {author} {\bibinfo {author} {\bibfnamefont {S.}~\bibnamefont
  {Endo}},\ }\emph {\bibinfo {title} {Hybrid quantum-classical algorithms and
  error mitigation}},\ \href@noop {} {Ph.D. thesis},\ \bibinfo  {school}
  {University of Oxford} (\bibinfo {year} {2019})\BibitemShut {NoStop}%
\bibitem [{\citenamefont {Endo}\ \emph
  {et~al.}(2020{\natexlab{b}})\citenamefont {Endo}, \citenamefont {Kurata},\
  and\ \citenamefont {Nakagawa}}]{PhysRevResearch.2.033281}%
  \BibitemOpen
  \bibfield  {author} {\bibinfo {author} {\bibfnamefont {S.}~\bibnamefont
  {Endo}}, \bibinfo {author} {\bibfnamefont {I.}~\bibnamefont {Kurata}}, \ and\
  \bibinfo {author} {\bibfnamefont {Y.~O.}\ \bibnamefont {Nakagawa}},\ }\href
  {\doibase 10.1103/PhysRevResearch.2.033281} {\bibfield  {journal} {\bibinfo
  {journal} {Phys. Rev. Research}\ }\textbf {\bibinfo {volume} {2}},\ \bibinfo
  {pages} {033281} (\bibinfo {year} {2020}{\natexlab{b}})}\BibitemShut
  {NoStop}%
\bibitem [{\citenamefont {Mitarai}\ and\ \citenamefont
  {Fujii}(2019)}]{mitarai2019methodology}%
  \BibitemOpen
  \bibfield  {author} {\bibinfo {author} {\bibfnamefont {K.}~\bibnamefont
  {Mitarai}}\ and\ \bibinfo {author} {\bibfnamefont {K.}~\bibnamefont
  {Fujii}},\ }\href@noop {} {\bibfield  {journal} {\bibinfo  {journal}
  {Physical Review Research}\ }\textbf {\bibinfo {volume} {1}},\ \bibinfo
  {pages} {013006} (\bibinfo {year} {2019})}\BibitemShut {NoStop}%
\bibitem [{\citenamefont {Li}\ \emph {et~al.}(2020)\citenamefont {Li},
  \citenamefont {Albash},\ and\ \citenamefont {Lidar}}]{Li_2020}%
  \BibitemOpen
  \bibfield  {author} {\bibinfo {author} {\bibfnamefont {R.~Y.}\ \bibnamefont
  {Li}}, \bibinfo {author} {\bibfnamefont {T.}~\bibnamefont {Albash}}, \ and\
  \bibinfo {author} {\bibfnamefont {D.~A.}\ \bibnamefont {Lidar}},\ }\href
  {\doibase 10.1088/2058-9565/ab9aab} {\bibfield  {journal} {\bibinfo
  {journal} {Quantum Science and Technology}\ }\textbf {\bibinfo {volume}
  {5}},\ \bibinfo {pages} {045010} (\bibinfo {year} {2020})}\BibitemShut
  {NoStop}%
\bibitem [{\citenamefont {Landau}\ and\ \citenamefont
  {Binder}(2014)}]{landau2014guide}%
  \BibitemOpen
  \bibfield  {author} {\bibinfo {author} {\bibfnamefont {D.~P.}\ \bibnamefont
  {Landau}}\ and\ \bibinfo {author} {\bibfnamefont {K.}~\bibnamefont
  {Binder}},\ }\href@noop {} {\emph {\bibinfo {title} {A guide to Monte Carlo
  simulations in statistical physics}}}\ (\bibinfo  {publisher} {Cambridge
  University Press},\ \bibinfo {year} {2014})\BibitemShut {NoStop}%
\bibitem [{\citenamefont {Sutskever}\ and\ \citenamefont
  {Tieleman}(2010)}]{sutskever2010convergence}%
  \BibitemOpen
  \bibfield  {author} {\bibinfo {author} {\bibfnamefont {I.}~\bibnamefont
  {Sutskever}}\ and\ \bibinfo {author} {\bibfnamefont {T.}~\bibnamefont
  {Tieleman}},\ }in\ \href@noop {} {\emph {\bibinfo {booktitle} {Proceedings of
  the thirteenth international conference on artificial intelligence and
  statistics}}}\ (\bibinfo {organization} {JMLR Workshop and Conference
  Proceedings},\ \bibinfo {year} {2010})\ pp.\ \bibinfo {pages}
  {789--795}\BibitemShut {NoStop}%
\bibitem [{\citenamefont {Cho}\ \emph {et~al.}(2010)\citenamefont {Cho},
  \citenamefont {Raiko},\ and\ \citenamefont {Ilin}}]{cho2010parallel}%
  \BibitemOpen
  \bibfield  {author} {\bibinfo {author} {\bibfnamefont {K.}~\bibnamefont
  {Cho}}, \bibinfo {author} {\bibfnamefont {T.}~\bibnamefont {Raiko}}, \ and\
  \bibinfo {author} {\bibfnamefont {A.}~\bibnamefont {Ilin}},\ }in\ \href@noop
  {} {\emph {\bibinfo {booktitle} {The 2010 international joint conference on
  neural networks (ijcnn)}}}\ (\bibinfo {organization} {IEEE},\ \bibinfo {year}
  {2010})\ pp.\ \bibinfo {pages} {1--8}\BibitemShut {NoStop}%
\bibitem [{\citenamefont {Desjardins}\ \emph {et~al.}(2010)\citenamefont
  {Desjardins}, \citenamefont {Courville}, \citenamefont {Bengio},
  \citenamefont {Vincent},\ and\ \citenamefont
  {Delalleau}}]{desjardins2010tempered}%
  \BibitemOpen
  \bibfield  {author} {\bibinfo {author} {\bibfnamefont {G.}~\bibnamefont
  {Desjardins}}, \bibinfo {author} {\bibfnamefont {A.}~\bibnamefont
  {Courville}}, \bibinfo {author} {\bibfnamefont {Y.}~\bibnamefont {Bengio}},
  \bibinfo {author} {\bibfnamefont {P.}~\bibnamefont {Vincent}}, \ and\
  \bibinfo {author} {\bibfnamefont {O.}~\bibnamefont {Delalleau}},\ }in\
  \href@noop {} {\emph {\bibinfo {booktitle} {Proceedings of the thirteenth
  international conference on artificial intelligence and statistics}}}\
  (\bibinfo {organization} {JMLR Workshop and Conference Proceedings},\
  \bibinfo {year} {2010})\ pp.\ \bibinfo {pages} {145--152}\BibitemShut
  {NoStop}%
\bibitem [{\citenamefont {Somma}\ \emph {et~al.}(2007)\citenamefont {Somma},
  \citenamefont {Batista},\ and\ \citenamefont {Ortiz}}]{somma2007quantum}%
  \BibitemOpen
  \bibfield  {author} {\bibinfo {author} {\bibfnamefont {R.~D.}\ \bibnamefont
  {Somma}}, \bibinfo {author} {\bibfnamefont {C.~D.}\ \bibnamefont {Batista}},
  \ and\ \bibinfo {author} {\bibfnamefont {G.}~\bibnamefont {Ortiz}},\
  }\href@noop {} {\bibfield  {journal} {\bibinfo  {journal} {Physical review
  letters}\ }\textbf {\bibinfo {volume} {99}},\ \bibinfo {pages} {030603}
  (\bibinfo {year} {2007})}\BibitemShut {NoStop}%
\bibitem [{\citenamefont {Yamamoto}\ \emph {et~al.}(2020)\citenamefont
  {Yamamoto}, \citenamefont {Ohzeki},\ and\ \citenamefont
  {Tanaka}}]{yamamoto2020fair}%
  \BibitemOpen
  \bibfield  {author} {\bibinfo {author} {\bibfnamefont {M.}~\bibnamefont
  {Yamamoto}}, \bibinfo {author} {\bibfnamefont {M.}~\bibnamefont {Ohzeki}}, \
  and\ \bibinfo {author} {\bibfnamefont {K.}~\bibnamefont {Tanaka}},\
  }\href@noop {} {\bibfield  {journal} {\bibinfo  {journal} {Journal of the
  Physical Society of Japan}\ }\textbf {\bibinfo {volume} {89}},\ \bibinfo
  {pages} {025002} (\bibinfo {year} {2020})}\BibitemShut {NoStop}%
\bibitem [{\citenamefont {Wu}\ and\ \citenamefont
  {Hsieh}(2019)}]{wu2019variational}%
  \BibitemOpen
  \bibfield  {author} {\bibinfo {author} {\bibfnamefont {J.}~\bibnamefont
  {Wu}}\ and\ \bibinfo {author} {\bibfnamefont {T.~H.}\ \bibnamefont {Hsieh}},\
  }\href@noop {} {\bibfield  {journal} {\bibinfo  {journal} {Physical Review
  Letters}\ }\textbf {\bibinfo {volume} {123}},\ \bibinfo {pages} {220502}
  (\bibinfo {year} {2019})}\BibitemShut {NoStop}%
\bibitem [{\citenamefont {Zhu}\ \emph {et~al.}(2019)\citenamefont {Zhu},
  \citenamefont {Johri}, \citenamefont {Linke}, \citenamefont {Landsman},
  \citenamefont {Nguyen}, \citenamefont {Alderete}, \citenamefont {Matsuura},
  \citenamefont {Hsieh},\ and\ \citenamefont {Monroe}}]{zhu2019variational}%
  \BibitemOpen
  \bibfield  {author} {\bibinfo {author} {\bibfnamefont {D.}~\bibnamefont
  {Zhu}}, \bibinfo {author} {\bibfnamefont {S.}~\bibnamefont {Johri}}, \bibinfo
  {author} {\bibfnamefont {N.}~\bibnamefont {Linke}}, \bibinfo {author}
  {\bibfnamefont {K.}~\bibnamefont {Landsman}}, \bibinfo {author}
  {\bibfnamefont {N.}~\bibnamefont {Nguyen}}, \bibinfo {author} {\bibfnamefont
  {C.}~\bibnamefont {Alderete}}, \bibinfo {author} {\bibfnamefont
  {A.}~\bibnamefont {Matsuura}}, \bibinfo {author} {\bibfnamefont
  {T.}~\bibnamefont {Hsieh}}, \ and\ \bibinfo {author} {\bibfnamefont
  {C.}~\bibnamefont {Monroe}},\ }\href@noop {} {\bibfield  {journal} {\bibinfo
  {journal} {arXiv preprint arXiv:1906.02699}\ } (\bibinfo {year}
  {2019})}\BibitemShut {NoStop}%
\bibitem [{\citenamefont {Verdon}\ \emph {et~al.}(2019)\citenamefont {Verdon},
  \citenamefont {Marks}, \citenamefont {Nanda}, \citenamefont {Leichenauer},\
  and\ \citenamefont {Hidary}}]{verdon2019quantum}%
  \BibitemOpen
  \bibfield  {author} {\bibinfo {author} {\bibfnamefont {G.}~\bibnamefont
  {Verdon}}, \bibinfo {author} {\bibfnamefont {J.}~\bibnamefont {Marks}},
  \bibinfo {author} {\bibfnamefont {S.}~\bibnamefont {Nanda}}, \bibinfo
  {author} {\bibfnamefont {S.}~\bibnamefont {Leichenauer}}, \ and\ \bibinfo
  {author} {\bibfnamefont {J.}~\bibnamefont {Hidary}},\ }\href@noop {}
  {\bibfield  {journal} {\bibinfo  {journal} {arXiv preprint arXiv:1910.02071}\
  } (\bibinfo {year} {2019})}\BibitemShut {NoStop}%
\bibitem [{\citenamefont {Chowdhury}\ \emph {et~al.}(2020)\citenamefont
  {Chowdhury}, \citenamefont {Low},\ and\ \citenamefont
  {Wiebe}}]{chowdhury2020variational}%
  \BibitemOpen
  \bibfield  {author} {\bibinfo {author} {\bibfnamefont {A.~N.}\ \bibnamefont
  {Chowdhury}}, \bibinfo {author} {\bibfnamefont {G.~H.}\ \bibnamefont {Low}},
  \ and\ \bibinfo {author} {\bibfnamefont {N.}~\bibnamefont {Wiebe}},\
  }\href@noop {} {\bibfield  {journal} {\bibinfo  {journal} {arXiv preprint
  arXiv:2002.00055}\ } (\bibinfo {year} {2020})}\BibitemShut {NoStop}%
\bibitem [{\citenamefont {Wang}\ \emph {et~al.}(2020)\citenamefont {Wang},
  \citenamefont {Li},\ and\ \citenamefont {Wang}}]{wang2020variational}%
  \BibitemOpen
  \bibfield  {author} {\bibinfo {author} {\bibfnamefont {Y.}~\bibnamefont
  {Wang}}, \bibinfo {author} {\bibfnamefont {G.}~\bibnamefont {Li}}, \ and\
  \bibinfo {author} {\bibfnamefont {X.}~\bibnamefont {Wang}},\ }\href@noop {}
  {\bibfield  {journal} {\bibinfo  {journal} {arXiv preprint arXiv:2005.08797}\
  } (\bibinfo {year} {2020})}\BibitemShut {NoStop}%
\bibitem [{\citenamefont {Verdon}\ \emph {et~al.}(2017)\citenamefont {Verdon},
  \citenamefont {Broughton},\ and\ \citenamefont
  {Biamonte}}]{verdon2017quantum}%
  \BibitemOpen
  \bibfield  {author} {\bibinfo {author} {\bibfnamefont {G.}~\bibnamefont
  {Verdon}}, \bibinfo {author} {\bibfnamefont {M.}~\bibnamefont {Broughton}}, \
  and\ \bibinfo {author} {\bibfnamefont {J.}~\bibnamefont {Biamonte}},\
  }\href@noop {} {\bibfield  {journal} {\bibinfo  {journal} {arXiv preprint
  arXiv:1712.05304}\ } (\bibinfo {year} {2017})}\BibitemShut {NoStop}%
\bibitem [{\citenamefont {Broadbent}\ \emph {et~al.}(2009)\citenamefont
  {Broadbent}, \citenamefont {Fitzsimons},\ and\ \citenamefont
  {Kashefi}}]{broadbent2009universal}%
  \BibitemOpen
  \bibfield  {author} {\bibinfo {author} {\bibfnamefont {A.}~\bibnamefont
  {Broadbent}}, \bibinfo {author} {\bibfnamefont {J.}~\bibnamefont
  {Fitzsimons}}, \ and\ \bibinfo {author} {\bibfnamefont {E.}~\bibnamefont
  {Kashefi}},\ }in\ \href@noop {} {\emph {\bibinfo {booktitle} {2009 50th
  Annual IEEE Symposium on Foundations of Computer Science}}}\ (\bibinfo
  {organization} {IEEE},\ \bibinfo {year} {2009})\ pp.\ \bibinfo {pages}
  {517--526}\BibitemShut {NoStop}%
\bibitem [{\citenamefont {Morimae}\ and\ \citenamefont
  {Fujii}(2013)}]{morimae2013blind}%
  \BibitemOpen
  \bibfield  {author} {\bibinfo {author} {\bibfnamefont {T.}~\bibnamefont
  {Morimae}}\ and\ \bibinfo {author} {\bibfnamefont {K.}~\bibnamefont
  {Fujii}},\ }\href@noop {} {\bibfield  {journal} {\bibinfo  {journal}
  {Physical Review A}\ }\textbf {\bibinfo {volume} {87}},\ \bibinfo {pages}
  {050301} (\bibinfo {year} {2013})}\BibitemShut {NoStop}%
\bibitem [{\citenamefont {Kashefi}\ \emph {et~al.}(2020)\citenamefont
  {Kashefi}, \citenamefont {Leichtle}, \citenamefont {Music},\ and\
  \citenamefont {Ollivier}}]{kashefi2020securing}%
  \BibitemOpen
  \bibfield  {author} {\bibinfo {author} {\bibfnamefont {E.}~\bibnamefont
  {Kashefi}}, \bibinfo {author} {\bibfnamefont {D.}~\bibnamefont {Leichtle}},
  \bibinfo {author} {\bibfnamefont {L.}~\bibnamefont {Music}}, \ and\ \bibinfo
  {author} {\bibfnamefont {H.}~\bibnamefont {Ollivier}},\ }\href@noop {}
  {\bibfield  {journal} {\bibinfo  {journal} {arXiv preprint arXiv:2011.10005}\
  } (\bibinfo {year} {2020})}\BibitemShut {NoStop}%
\bibitem [{\citenamefont {et~al.}(2019)}]{Qiskit}%
  \BibitemOpen
  \bibfield  {author} {\bibinfo {author} {\bibfnamefont {H.~A.}\ \bibnamefont
  {et~al.}},\ }\href {\doibase 10.5281/zenodo.2562110} {\enquote {\bibinfo
  {title} {Qiskit: An open-source framework for quantum computing},}\ }
  (\bibinfo {year} {2019})\BibitemShut {NoStop}%
\bibitem [{\citenamefont {Zoufal}\ \emph {et~al.}(2020)\citenamefont {Zoufal},
  \citenamefont {Lucchi},\ and\ \citenamefont
  {Woerner}}]{zoufal2020variational}%
  \BibitemOpen
  \bibfield  {author} {\bibinfo {author} {\bibfnamefont {C.}~\bibnamefont
  {Zoufal}}, \bibinfo {author} {\bibfnamefont {A.}~\bibnamefont {Lucchi}}, \
  and\ \bibinfo {author} {\bibfnamefont {S.}~\bibnamefont {Woerner}},\
  }\href@noop {} {\bibfield  {journal} {\bibinfo  {journal} {arXiv preprint
  arXiv:2006.06004}\ } (\bibinfo {year} {2020})}\BibitemShut {NoStop}%
\bibitem [{\citenamefont {Temme}\ \emph {et~al.}(2017)\citenamefont {Temme},
  \citenamefont {Bravyi},\ and\ \citenamefont {Gambetta}}]{temme2017error}%
  \BibitemOpen
  \bibfield  {author} {\bibinfo {author} {\bibfnamefont {K.}~\bibnamefont
  {Temme}}, \bibinfo {author} {\bibfnamefont {S.}~\bibnamefont {Bravyi}}, \
  and\ \bibinfo {author} {\bibfnamefont {J.~M.}\ \bibnamefont {Gambetta}},\
  }\href@noop {} {\bibfield  {journal} {\bibinfo  {journal} {Physical review
  letters}\ }\textbf {\bibinfo {volume} {119}},\ \bibinfo {pages} {180509}
  (\bibinfo {year} {2017})}\BibitemShut {NoStop}%
\bibitem [{\citenamefont {Endo}\ \emph {et~al.}(2018)\citenamefont {Endo},
  \citenamefont {Benjamin},\ and\ \citenamefont {Li}}]{endo2018practical}%
  \BibitemOpen
  \bibfield  {author} {\bibinfo {author} {\bibfnamefont {S.}~\bibnamefont
  {Endo}}, \bibinfo {author} {\bibfnamefont {S.~C.}\ \bibnamefont {Benjamin}},
  \ and\ \bibinfo {author} {\bibfnamefont {Y.}~\bibnamefont {Li}},\ }\href@noop
  {} {\bibfield  {journal} {\bibinfo  {journal} {Physical Review X}\ }\textbf
  {\bibinfo {volume} {8}},\ \bibinfo {pages} {031027} (\bibinfo {year}
  {2018})}\BibitemShut {NoStop}%
\bibitem [{\citenamefont {Bonet-Monroig}\ \emph {et~al.}(2018)\citenamefont
  {Bonet-Monroig}, \citenamefont {Sagastizabal}, \citenamefont {Singh},\ and\
  \citenamefont {O'Brien}}]{bonet2018low}%
  \BibitemOpen
  \bibfield  {author} {\bibinfo {author} {\bibfnamefont {X.}~\bibnamefont
  {Bonet-Monroig}}, \bibinfo {author} {\bibfnamefont {R.}~\bibnamefont
  {Sagastizabal}}, \bibinfo {author} {\bibfnamefont {M.}~\bibnamefont {Singh}},
  \ and\ \bibinfo {author} {\bibfnamefont {T.}~\bibnamefont {O'Brien}},\
  }\href@noop {} {\bibfield  {journal} {\bibinfo  {journal} {Physical Review
  A}\ }\textbf {\bibinfo {volume} {98}},\ \bibinfo {pages} {062339} (\bibinfo
  {year} {2018})}\BibitemShut {NoStop}%
\bibitem [{\citenamefont {Endo}\ \emph {et~al.}(2019)\citenamefont {Endo},
  \citenamefont {Zhao}, \citenamefont {Li}, \citenamefont {Benjamin},\ and\
  \citenamefont {Yuan}}]{endo2019mitigating}%
  \BibitemOpen
  \bibfield  {author} {\bibinfo {author} {\bibfnamefont {S.}~\bibnamefont
  {Endo}}, \bibinfo {author} {\bibfnamefont {Q.}~\bibnamefont {Zhao}}, \bibinfo
  {author} {\bibfnamefont {Y.}~\bibnamefont {Li}}, \bibinfo {author}
  {\bibfnamefont {S.}~\bibnamefont {Benjamin}}, \ and\ \bibinfo {author}
  {\bibfnamefont {X.}~\bibnamefont {Yuan}},\ }\href@noop {} {\bibfield
  {journal} {\bibinfo  {journal} {Physical Review A}\ }\textbf {\bibinfo
  {volume} {99}},\ \bibinfo {pages} {012334} (\bibinfo {year}
  {2019})}\BibitemShut {NoStop}%
\bibitem [{\citenamefont {McArdle}\ \emph
  {et~al.}(2019{\natexlab{b}})\citenamefont {McArdle}, \citenamefont {Yuan},\
  and\ \citenamefont {Benjamin}}]{mcardle2019error}%
  \BibitemOpen
  \bibfield  {author} {\bibinfo {author} {\bibfnamefont {S.}~\bibnamefont
  {McArdle}}, \bibinfo {author} {\bibfnamefont {X.}~\bibnamefont {Yuan}}, \
  and\ \bibinfo {author} {\bibfnamefont {S.}~\bibnamefont {Benjamin}},\
  }\href@noop {} {\bibfield  {journal} {\bibinfo  {journal} {Physical review
  letters}\ }\textbf {\bibinfo {volume} {122}},\ \bibinfo {pages} {180501}
  (\bibinfo {year} {2019}{\natexlab{b}})}\BibitemShut {NoStop}%
\bibitem [{\citenamefont {Song}\ \emph {et~al.}(2019)\citenamefont {Song},
  \citenamefont {Cui}, \citenamefont {Wang}, \citenamefont {Hao}, \citenamefont
  {Feng},\ and\ \citenamefont {Li}}]{song2019quantum}%
  \BibitemOpen
  \bibfield  {author} {\bibinfo {author} {\bibfnamefont {C.}~\bibnamefont
  {Song}}, \bibinfo {author} {\bibfnamefont {J.}~\bibnamefont {Cui}}, \bibinfo
  {author} {\bibfnamefont {H.}~\bibnamefont {Wang}}, \bibinfo {author}
  {\bibfnamefont {J.}~\bibnamefont {Hao}}, \bibinfo {author} {\bibfnamefont
  {H.}~\bibnamefont {Feng}}, \ and\ \bibinfo {author} {\bibfnamefont
  {Y.}~\bibnamefont {Li}},\ }\href@noop {} {\bibfield  {journal} {\bibinfo
  {journal} {Science advances}\ }\textbf {\bibinfo {volume} {5}},\ \bibinfo
  {pages} {eaaw5686} (\bibinfo {year} {2019})}\BibitemShut {NoStop}%
\bibitem [{\citenamefont {Sagastizabal}\ \emph {et~al.}(2019)\citenamefont
  {Sagastizabal}, \citenamefont {Bonet-Monroig}, \citenamefont {Singh},
  \citenamefont {Rol}, \citenamefont {Bultink}, \citenamefont {Fu},
  \citenamefont {Price}, \citenamefont {Ostroukh}, \citenamefont
  {Muthusubramanian}, \citenamefont {Bruno} \emph
  {et~al.}}]{sagastizabal2019experimental}%
  \BibitemOpen
  \bibfield  {author} {\bibinfo {author} {\bibfnamefont {R.}~\bibnamefont
  {Sagastizabal}}, \bibinfo {author} {\bibfnamefont {X.}~\bibnamefont
  {Bonet-Monroig}}, \bibinfo {author} {\bibfnamefont {M.}~\bibnamefont
  {Singh}}, \bibinfo {author} {\bibfnamefont {M.~A.}\ \bibnamefont {Rol}},
  \bibinfo {author} {\bibfnamefont {C.}~\bibnamefont {Bultink}}, \bibinfo
  {author} {\bibfnamefont {X.}~\bibnamefont {Fu}}, \bibinfo {author}
  {\bibfnamefont {C.}~\bibnamefont {Price}}, \bibinfo {author} {\bibfnamefont
  {V.}~\bibnamefont {Ostroukh}}, \bibinfo {author} {\bibfnamefont
  {N.}~\bibnamefont {Muthusubramanian}}, \bibinfo {author} {\bibfnamefont
  {A.}~\bibnamefont {Bruno}},  \emph {et~al.},\ }\href@noop {} {\bibfield
  {journal} {\bibinfo  {journal} {Physical Review A}\ }\textbf {\bibinfo
  {volume} {100}},\ \bibinfo {pages} {010302} (\bibinfo {year}
  {2019})}\BibitemShut {NoStop}%
\bibitem [{\citenamefont {Kandala}\ \emph {et~al.}(2019)\citenamefont
  {Kandala}, \citenamefont {Temme}, \citenamefont {C{\'o}rcoles}, \citenamefont
  {Mezzacapo}, \citenamefont {Chow},\ and\ \citenamefont
  {Gambetta}}]{kandala2019error}%
  \BibitemOpen
  \bibfield  {author} {\bibinfo {author} {\bibfnamefont {A.}~\bibnamefont
  {Kandala}}, \bibinfo {author} {\bibfnamefont {K.}~\bibnamefont {Temme}},
  \bibinfo {author} {\bibfnamefont {A.~D.}\ \bibnamefont {C{\'o}rcoles}},
  \bibinfo {author} {\bibfnamefont {A.}~\bibnamefont {Mezzacapo}}, \bibinfo
  {author} {\bibfnamefont {J.~M.}\ \bibnamefont {Chow}}, \ and\ \bibinfo
  {author} {\bibfnamefont {J.~M.}\ \bibnamefont {Gambetta}},\ }\href@noop {}
  {\bibfield  {journal} {\bibinfo  {journal} {Nature}\ }\textbf {\bibinfo
  {volume} {567}},\ \bibinfo {pages} {491} (\bibinfo {year}
  {2019})}\BibitemShut {NoStop}%
\bibitem [{\citenamefont {Zhang}\ \emph {et~al.}(2020)\citenamefont {Zhang},
  \citenamefont {Lu}, \citenamefont {Zhang}, \citenamefont {Chen},
  \citenamefont {Li}, \citenamefont {Zhang},\ and\ \citenamefont
  {Kim}}]{zhang2020error}%
  \BibitemOpen
  \bibfield  {author} {\bibinfo {author} {\bibfnamefont {S.}~\bibnamefont
  {Zhang}}, \bibinfo {author} {\bibfnamefont {Y.}~\bibnamefont {Lu}}, \bibinfo
  {author} {\bibfnamefont {K.}~\bibnamefont {Zhang}}, \bibinfo {author}
  {\bibfnamefont {W.}~\bibnamefont {Chen}}, \bibinfo {author} {\bibfnamefont
  {Y.}~\bibnamefont {Li}}, \bibinfo {author} {\bibfnamefont {J.-N.}\
  \bibnamefont {Zhang}}, \ and\ \bibinfo {author} {\bibfnamefont
  {K.}~\bibnamefont {Kim}},\ }\href@noop {} {\bibfield  {journal} {\bibinfo
  {journal} {Nature communications}\ }\textbf {\bibinfo {volume} {11}},\
  \bibinfo {pages} {1} (\bibinfo {year} {2020})}\BibitemShut {NoStop}%
\bibitem [{\citenamefont {Hakoshima}\ \emph {et~al.}(2020)\citenamefont
  {Hakoshima}, \citenamefont {Matsuzaki},\ and\ \citenamefont
  {Endo}}]{hakoshima2020relationship}%
  \BibitemOpen
  \bibfield  {author} {\bibinfo {author} {\bibfnamefont {H.}~\bibnamefont
  {Hakoshima}}, \bibinfo {author} {\bibfnamefont {Y.}~\bibnamefont
  {Matsuzaki}}, \ and\ \bibinfo {author} {\bibfnamefont {S.}~\bibnamefont
  {Endo}},\ }\href@noop {} {\bibfield  {journal} {\bibinfo  {journal} {arXiv
  preprint arXiv:2009.12759}\ } (\bibinfo {year} {2020})}\BibitemShut {NoStop}%
\end{thebibliography}%
\end{document}